\documentclass[longauth]{aa} 
\usepackage{colortbl}
\usepackage{rotating}
\usepackage{caption}
\usepackage{tabularx}
\usepackage{graphicx}
\usepackage{txfonts}
\usepackage{comment}
\usepackage{float}
\usepackage{multirow}
\usepackage{subcaption}
\usepackage[colorlinks=true, linkcolor=blue, citecolor=blue]{hyperref}
\usepackage[nameinlink,noabbrev]{cleveref}
\usepackage{fancyhdr}
\usepackage{bm}
\usepackage{xcolor}
\begin{document} 

   \title{J-PAS $\&$ FLAMINGO: Cosmic voids and void galaxies in the gravitational landscape of photometric surveys}

   \author{J.A. Mansour\thanks{\email{jad-alexandru.mansour@ut.ee}}$^1$, B. McCarthy$^{2}$, L. J. Liivamägi$^1$, A. Tamm$^1$, R. van de Weygaert$^{2}$, J. Laur$^1$, E. Tempel$^{1}$, M. Einasto$^1$, P. Heinämäki$^{3}$, J. Schaye$^{4}$, M. Schaller$^{4, 5}$, R. Abramo$^{6, 7}$,  A. Hernán-Caballero$^8$, V. Marra$^{18, 19, 20}$, J. Alcaniz$^{9}$, N. Benitez$^{21}$, S. Bonoli$^{8,11}$, S. Carneiro$^{9}$, J. Cenarro$^{8}$, D. Cristóbal-Hornillos$^{8}$, S. Daflon$^{9}$ , R. Dupke$^{9, 10, 13}$, A. Ederoclite$^{8}$,  Rosa M. González Delgado$^{10}$ , C. Hernández-Monteagudo$^{14, 15}$ , J. Liu$^{16}$ , C. López-Sanjuan$^{8}$, A. Marín-Franch$^{8}$, C. M. de Oliveira$^{7}$, M. Moles$^{8, 10}$, F. Roig$^{9}$, L. Sodré Jr$^{7}$, K. Taylor$^{17}$, J. Varela$^{8}$, H. Vázquez Ramió$^{8}$, J. Vilchez$^{10}$, J. Zaragoza-Cardiel$^{8}$
          }
   \authorrunning{Short Author List}
   \institute{
   $^{1}$  Tartu Observatory, University of Tartu, Observatooriumi 1, 61602 Tõravere, Estonia \\
   $^{2}$ Kapteyn Astronomical Institute, University of Groningen, PO Box 800, NL-9747 AD, Groningen, the Netherlands \\
   $^{3}$ Tuorla Observatory, Department of Physics and Astronomy, Vesilinnantie 5, University of Turku, 20014 Turku, Finland \\
   $^{4}$ Leiden Observatory, Leiden University, PO Box 9513, NL-2300 RA Leiden, the Netherlands \\
   $^{5}$ Lorentz Institute for Theoretical Physics, Leiden University, PO box 9506, NL-2300 RA Leiden, the Netherlands \\
   $^{6}$Instituto de Física, Universidade de São Paulo, Rua do Matão 1371, 05508-090 São Paulo, Brazil \\
   $^{7}$ Departamento de Astronomia, Instituto de Astronomia, Geofísica
e Ciências Atmosféricas da USP, Cidade Universitária, 05508-900,
São Paulo, SP, Brazil \\
    $^8$ Centro de Estudios de Física del Cosmos de Aragón (CEFCA), Plaza San Juan, 1, E-44001 Teruel, Spain \\
    $^{9}$ Observatório Nacional, Ministério da Ciência, Tecnologia, Inovação e Comunicações, Rua General José Cristino, 77, São Cristóvão, 20921-400, Rio de Janeiro, Brazil \\
    $^{10}$ Instituto de Astrofísica de Andalucía (CSIC), P.O. Box 3004, 18080 Granada, Spain \\
        $^{11}$ Donostia International Physics Centre, Paseo Manuel de Lardizabal 4, 20018 Donostia-San Sebastian, Spain \\
    $^{12}$ Instituto de Física, Universidade Federal da Bahia, 40210-340,
Salvador, BA, Brazil \\
$^{13}$  Department of Astronomy, University of Michigan, 311 West Hall,
1085 South University Ave., Ann Arbor, USA \\
$^{14}$ Departamento de Astrofísica, Universidad de La Laguna, 38206,
La Laguna, Tenerife, Spain \\
$^{15}$ Instituto de Astrofísica de Canarias, 38200 La Laguna, Tenerife,
Spain       \\
$^{16}$ Chinese National Astronomical Observatory of China,
Chinese Academy of Sciences, Beijing, China 
$^{17}$ Instruments4, 4121 Pembury Place, La Cañada-Flintridge, CA 91011,
USA \\ 
$^{18} $ Departamento de Física, Universidade Federal de Ouro Preto, 35400-000, Ouro Preto, MG, Brazil \\
$^{19}$  INAF – Osservatorio Astronomico di Trieste, via Tiepolo 11, 34131
Trieste, Italy \\
$^{20}$ IFPU – Institute for Fundamental Physics of the Universe, via
Beirut 2, 34151 Trieste, Italy \\
$^{21}$ Independent Researcher
}

   \date{}

  \abstract  
    {Photometric surveys offer a powerful way to map the large-scale structure of the Universe. Cosmic voids are an essential component of this structure, but their identification in photometric surveys is complicated by the redshift errors of galaxy tracers. This poses a significant challenge in investigating the environmental effect of cosmic voids on galaxy properties.}
    {We provide a \textit{two-tier approach} to robustly identify and characterise dynamically relevant cosmic voids and their resident galaxies in galaxy mocks of the Javalambre Physics of the Accelerating Universe Astrophysical Survey (J-PAS), assessing whether established trends in void galaxy properties can be detected despite photometric redshift errors.}   
    {Using FLAMINGO simulation galaxy mocks at $z=0.3$ and $m_i < 20$, we compare a FLAMINGO-based ideal (\textit{FBI}) sample to a FLAMINGO-based \textit{JP} sample incorporating J-PAS-like photometric redshift errors. We mitigate these errors via a quasi-gravitational potential field, computed from the logarithm of the galaxy number density, within the two galaxy mocks. We apply a watershed algorithm to a thresholded quasi-potential field, allowing us to robustly identify dynamically dominant voids within the simulation volume. Subsequently, based on the density field, we define a sample of massive void galaxies ($M_*$ $\geq$ 10$^{10}$ $M_{\odot}$) and a comparison sample of galaxies located in high-density regions.}
    {We evaluate our approach by comparing the void population identified in the \textit{FBI} sample against that from the \textit{JP} sample. While photometric errors lead to a slightly lower void abundance and a marginal shift toward larger, less spherical shapes, the overall size and ellipticity distributions show a strong correspondence between the two mocks. The primary impact of photometric uncertainties manifests in the density profiles of the \textit{JP} sample voids as a contamination of void interiors with galaxies scattered from the high-density regions. Moreover, we recover a reasonable number of individual voids in the \textit{FBI} sample from the \textit{JP} sample. These recovered voids exhibit excellent agreement in their size and shape distributions, while collectively occupying approximately $63\%$ of the thresholded quasi-potential volume. Furthermore, we find that in both galaxy mocks, void galaxies tend to have lower stellar masses, bluer colours, and exhibit enhanced star formation activity compared to galaxies of the same stellar mass in high-density environments.}
    {Our findings suggest that a quasi-gravitational potential may serve as an effective tool for mitigating redshift errors of the order expected for the J-PAS survey. Taking advantage of the quasi-potential smoothing, we identify and characterise reliable, dynamically dominant underdensities that are less sensitive to small-scale noise. Furthermore, the population of massive void galaxies exhibits expected trends in their properties when compared to galaxies located in higher-density environments.}
    
    {}

    
   \maketitle





\section{Introduction}
The current study investigates the implications of using photometric redshift data for characterising properties of voids, such as size and
shape, as well as the properties of the void galaxies residing in them. While photometric surveys offer an efficient means of mapping the distribution of a large number of galaxies and probing their intrinsic properties over deep redshift ranges, the lower accuracy of photometric redshifts introduces substantial distortions, blurring the underlying cosmic structure. This will strongly affect the identification of voids, the inferred void properties, the detectability of small voids,
void substructure and, related to this, the ability to extract an unbiased inventory of void galaxy properties. Here we investigate the use of a dynamics-based void detection formalism in alleviating the effects of the sizeable photometric redshift errors.

The large-scale structure of the Universe resembles a web-like pattern, the {\it Cosmic Web}, which consists of filaments, clusters, and walls that delineate vast, low-density regions known as cosmic voids \citep{1970A&A.....5...84Z, 1978MNRAS.185..357J, 1986ApJ...302L...1D, 1989RvMP...61..185S, 1996Natur.380..603B, 2008LNP...740..409V,2010MNRAS.408.2163A, 2014MNRAS.441.2923C}. Filaments are the most visually outstanding features of the large-scale Universe, in which around $50\%$ of the mass in the Universe resides. On the other hand, almost 80\% of the cosmic volume belongs to the interior of voids~\citep[see e.g.][]{2009MNRAS.396.1815F,2014MNRAS.441.2923C,Ganeshaiah_Veena_2019, 2024ApJ...962...58C}. Voids are enormous regions with sizes in the range of $20-50h^{-1}$ Mpc that are practically devoid of any galaxies, and which occupy a major share of space in the Universe. Forming an essential and prominent aspect of the {\it Cosmic Web}, their dynamical impact
is instrumental in the spatial organisation of the large-scale structure of the Universe  \citep{1984MNRAS.206P...1I,2004MNRAS.350..517S, 2026MNRAS.547ag193K}. For comprehensive reviews on their
nature see e.g. \cite{2011IJMPS...1...41V, 2016IAUS..308..493V}. Furthermore, voids serve as a fundamental cosmological probe, utilised to constrain cosmological parameters \citep{2010MNRAS.403.1392L,2012MNRAS.426..440B, 2014MNRAS.443.2983S, Hamaus_2016, 2024A&A...682A..20C}, detect baryon acoustic oscillations \citep[BAO;][]{Kitaura_2016, 2023MNRAS.526.2889T}, and place limits on the neutrino mass \citep{Schuster_2019, 10.1093/mnras/stz1944}.

Cosmic voids have been a subject of observational study for nearly half a century, with early detections dating back to the late 1970s \citep{1978MNRAS.185..357J,1978ApJ...222..784G,1982Natur.300..407Z,1987ApJ...314..493K}. Their prominence throughout the spatial galaxy distribution was confirmed and firmly established by a range of large redshift surveys, first by the Center for Astrophysics survey \citep[CfA;][]{1986ApJ...302L...1D} and soon thereafter by the large wide-field maps obtained by the 2dFGRS, SDSS and 2MRS surveys \citep[e.g.,][]{colless20032df, 2012ApJS..199...26H}. The development of solidly defined void detection techniques, such as those involving topological watershed formalisms \citep{2007MNRAS.380..551P,2008MNRAS.386.2101N,2015A&C.....9....1S}, enabled the extraction of void catalogues from large galaxy redshift surveys \citep[e.g.][]{2012ApJ...761...44S, 2014MNRAS.440.1248N}. 
Modern large-scale structure studies rely on high-precision data, which has ushered in a wave of precision cosmology studies \citep[for comprehensive reviews of their cosmological impact, see e.g.][]{2019BAAS...51c..40P, 2026arXiv260114362C}. It will be marked by an unprecedented avalanche of data produced by ongoing and forthcoming surveys, including Euclid \citep{2025A&A...697A...1E}, DESI \citep{2022AJ....164..207D}, 4MOST \citep{2019Msngr.175....3D}, and the Vera C. Rubin Observatory’s LSST \citep{2019ApJ...873..111I}. It will allow the reconstruction of the detailed cosmic web's topological landscape, a prerequisite for the accurate identification and analysis of cosmic voids.

A fundamental aspect of voids is that their identity is defined in relation to the neighbouring structures \citep{2007MNRAS.380..551P,2011IJMPS...1...41V, 2016IAUS..308..493V}. This aspect also implies that the evolution of voids results from the intricate interaction with their surroundings. As voids expand, they collide and interact with surrounding structures. As a result, they tend to merge with surrounding peers, yielding a population of ever larger voids emerging through the implied hierarchical evolutionary sequence
\citep{1993MNRAS.263..481V, 1993ApJ...410..458D, 2004MNRAS.350..517S,2013MNRAS.428.3409A, 2014MNRAS.440.1248N,2026MNRAS.547ag193K}. The remnants of this process are still visible in the multiscale substructure of voids \citep{2013MNRAS.428.3409A, 2016IAUS..308..493V}. Moreover, because of their naturally modest density deficit, their dynamical evolution remains strongly influenced by the gravitational forces exerted by surrounding filaments, walls, and voids \citep{2026MNRAS.547ag193K}. External gravitational forces and
tides may even evoke a contraction of voids along one or more directions, which is often seen to be happening to subvoids near the boundaries of larger expanding voids. If embedded in a larger-scale overdensity, they may even collapse out of existence due to the inward gravitational contraction of the overdensity. This \textit{void-in-cloud} process is responsible for a peaked
\textit{Void Size Distribution}, yielding a characteristic void size whose value is dictated by the primordial spectrum of density and velocity perturbations, by the cosmological era, and by a range of cosmological parameters \citep{2004MNRAS.350..517S,2013MNRAS.434.2167J,2024JCAP...10..079V}. 

A major point of interest is the population of galaxies located within voids, the {\it void galaxies}. The existence of void galaxies ties in with the important question of the role of cosmic environment in forming and shaping the structure and properties of galaxies \citep[for a recent theoretical and systematic study see][]{2026arXiv260418209H}. Since the seminal work of \cite{1980ApJ...236..351D}, it is known that there is a strong relation between the morphology of galaxies and the density of the
cosmic environment in which they are embedded. Elliptical galaxies tend to reside in high-density regions of clusters, while spiral galaxies prefer more moderate and sparser regions. Within this context, the pristine environment of voids represents an ideal and pure setting for the study of galaxy formation. Largely unaffected by the complex processes shaping the galaxies in high-density environments, the isolated void regions must hold important clues to the formation and evolution of galaxies. Several observational programs have been exploring the characteristics of void galaxies. Following the systematic SDSS-based study of void galaxies by Vogeley and collaborators
\citep{2004ApJ...617...50R, 2005ApJ...620..618H,2005ApJ...624..571R,2012MNRAS.426.3041H}, the multiwavelength Void Galaxy Survey (VGS) has been investigating the gas content, star formation activity and structural parameters of galaxies living in voids \citep{2011AJ....141....4K,Kreckel_2012,2017MNRAS.464..666B}. Building on SDSS data within the CAVITY framework, \cite{2023Natur.619..269D} provided detailed constraints on the star formation histories of void galaxies. The ongoing CAVITY survey \citep{2024A&A...687A..98C,2024A&A...689A.213P, 2024A&A...691A.161G} extends this effort with a more comprehensive program that includes IFU spectroscopic observations. From previous studies, we know that nearly all void galaxies are faint, late-type, often somewhat irregular, with blue colours \citep{1991MNRAS.250..802E, Grogin_1999, Kreckel_2012, 2012MNRAS.426.3041H, 2021ApJ...906...97F, Rodr_guez_Medrano_2023,  2024ApJ...962...58C, 2026arXiv260418209H}, gas-rich \citep{Kreckel_2012, 2021ApJ...906...97F, 2022MNRAS.517..712R, 2026A&A...709A.227C}, have a high star formation rate (SFR) \citep{Kreckel_2012, 2017MNRAS.464..666B, 2021ApJ...906...97F, Rodr_guez_Medrano_2023, 2026arXiv260418209H}, and low stellar masses \citep{Kreckel_2012, 2021ApJ...906...97F,  2020MNRAS.493..899H,  2022A&A...668A..69E,  2024ApJ...962...58C}.



So far, voids have been predominantly identified in galaxy spectroscopic redshift catalogues, as photometric redshift errors are significant and smear the line-of-sight positions of galaxies. In our previous work \citep{2025A&A...695A.174M}, we demonstrated that the direct identification of voids in a photometric mock catalogue is heavily affected by redshift uncertainties; specifically, void properties show significant deviations when compared to an error-free reference catalogue. Studies addressing this specific challenge remain sparse. An option put forward by  \cite{2017MNRAS.465..746S} is to mitigate the uncertainties by projecting galaxies into 2D slices, and subsequently identifying the voids in the projected density field of the slice. The suggested conclusion is that the voids are best recovered when the slice has a thickness comparable to the photo-$z$ uncertainty. In combating the effect of photometric redshift errors, a recent paper by \cite{2025MNRAS.538.2050H} showed that reconstruction of the density field via Poisson processes is feasible when both photometric and spectroscopic data are available.

To analyse the void population and void galaxies in photo-$z$ survey maps, we focus on the gravitational potential field rather than the density field. This approach allows us to isolate only the dynamically dominant voids during identification. It involves the recognition that the large expanding voids that represent the true expansion centres in the cosmic matter distribution should be well identifiable even in the case of major redshift uncertainties such as those encountered in photo-$z$ surveys. Smaller voids, either subvoids of these major expansion regions or the ones that are gravitationally dominated by their peers or surrounding filaments and walls
\citep[see][for a thorough discussion]{2026MNRAS.547ag193K}, are not considered in this void identification.

To identify voids within the gravitational potential field, we invoke the
watershed transform, delineating the region over which the expanding voids hold sway \citep{theses_fse8301}. Many studies of voids are based on the use of watershed-based formalisms to outline the regions around the minima in the matter or galaxy density field. Introduced and proposed by \cite{theses_fse8301, 2007MNRAS.380..551P} identifies voids via a watershed transform applied to the Delaunay Tessellation Field Estimator (DTFE) density field reconstruction \citep{2000A&A...363L..29S, 2009LNP...665..291V,2011MNRAS.416.2494P}. It delineates the region of a void independent of its scale and shape, and provides a direct objective measurement of the volume, shape and orientation of the voids. Since its introduction, watershed void identification has proliferated in the development of a range of different versions that involve variations
in probabilistic and technical details \citep[e.g.][]{2008MNRAS.387..933C,2008MNRAS.386.2101N,2015A&C.....9....1S}.

 For the watershed identification of the gravitationally dominant voids, we outline the descending gravitational potential
manifold around peaks in the gravitational potential field. It is
found that these supervoids are quite well identified in photo-z maps, i.e. their properties correspond well to the same voids
in the underlying physical mass distribution.  The gravitational
void watershed formalism \citep{Mccarthy2024} follows a philosophy similar to
that of the watershed identification of large supercluster complexes, such as Laniakea, with the attraction basins in the gravitational potential \citep[see][]{2014Natur.513...71T,2020MNRAS.493.3513D,2023A&A...678A.176D}, parallel to the original watershed
application to the density field \citep{2007MNRAS.380..551P}.

For the study of the massive void galaxies, we follow up the void identification with tracing the void galaxies within a spherical volume around the void centre. For a comparison sample, we additionally select galaxies residing in high-density regions. In the present study, we compare the properties of void populations identified in a baseline galaxy mock sample against those in a mock sample incorporating realistic J-PAS redshift uncertainties. Amongst others, we investigate whether for the identified void galaxies, familiar trends in galaxy properties (colour, stellar mass, and star formation activity - SFR and sSFR) can be observed despite redshift errors when compared to a sample of galaxies residing in high-density regions.

This paper is organised as follows. In Section~\ref{sec:data}, we describe the J-PAS survey, the J-PAS galaxy mocks (the comparison mock, \textit{FBI}, and the mock with realistic J-PAS photometric redshift errors, \textit{JP}) based on FLAMINGO hydrodynamical simulation and the modelling of redshift errors. Section~\ref{sec:meth} outlines the computation of the galaxy number density and quasi-gravitational potential fields from our galaxy mocks, alongside the watershed-based identification of dynamically relevant voids. We further describe the derivation of void physical properties -- specifically sizes, shapes, and density profiles -- and define the selection criteria for both void galaxies and a high-density comparison sample. Our results start from Section~\ref{sec:void_results}, where we present a characterisation of void properties. We then compare these properties between our two J-PAS galaxy mocks to assess their similarity, followed by a brief evaluation of the void recovery. In Section~\ref{sec: void galaxies}, we provide a comparative analysis of galaxies in voids versus high-density environments, examining their stellar masses, star formation activities, and colours. We evaluate whether the expected environmental trends are preserved in the \textit{JP} mock sample, with respect to the \textit{FBI}. Finally, Section~\ref{sec:conc} concludes with a summary of our work, followed by a discussion on limitations and potential application to future surveys.

\section{Data: FLAMINGO simulation and J-PAS mocks}
\label{sec:data}


\subsection{The J-PAS survey}

The J-PAS survey \citep{2014arXiv1403.5237B, 2021A&A...653A..31B} is an ongoing photometric survey conducted at the Observatorio Astrofísico de Javalambre (OAJ) and developed by the Centro de Estudios de Física del Cosmos de Aragón (CEFCA). It employs the Javalambre Survey Telescope (JST/T250), a Ritchey-Chrétien telescope with a 2.55~m aperture and a 3-degree-diameter field of view field of view. J-PAS distinguishes itself from other surveys by its unique filter system, which consists of 54 narrow-band filters, two intermediate-band filters, and one broad-band filter, covering a wavelength range from 3780~\AA\ to 9100~\AA\ and providing a low-resolution spectrum (J-spectra) for each observed pixel. The J-PAS survey was designed to observe an area of approximately 8500~deg$^{2}$ of the Northern Sky and is expected to provide photometric redshifts for $\sim10^{8}$ luminous red and emission-line galaxies, several million quasi-stellar objects, and roughly $7\times 10^{5}$ galaxy groups and clusters up to $z \sim 1.3$.

Multiple methods have been developed for estimating photometric redshifts \citep[for an overview of methods, see][]{2022ARA&A..60..363N}. The J-PAS survey capabilities for precise photometric redshifts have been demonstrated in a series of works \citep{2021A&A...654A.101H, 2023A&A...671A..71H, 2022A&A...668A...8L}, along with its potential for large-scale structure studies. Early results include analyses of galaxy populations \citep{2021A&A...649A..79G,2021A&A...647A.158M,2022A&A...666A..84G}, the detection and characterisation of galaxy groups and clusters using the AMICO algorithm \citep{2023A&A...678A.145M, 2024A&A...685A..98D}, and the identification of quasars with accurate redshift estimation \citep{2023A&A...673A.103M, 2023A&A...678A.144P}. 

In this work, we made use of the TOPz pipeline \citep{2022A&A...668A...8L, 2025A&A...703A.261T}, which provides template-based photo-$z$ estimates with J-PAS specific features. The TOPz pipeline has already produced redshifts for the miniJPAS, which fulfilled the expectation for the J-PAS: $38.6\%$ of galaxies with $m_{\rm r} < 22$ mag have reached the redshift accuracy goal $dz/(1+z) < 0.003$, where the redshift errors are calculated as :    
\begin{align}\hspace{0.3cm}
    z_{\rm err} = \frac{z_{\rm phot} - z_{\rm spec}}{1 + z_{\rm spec}},
    \label{zerr}
\end{align}
where $z_{\rm phot}$ is the estimated photometric redshift while $z_{\rm spec}$ is the spectroscopic redshift, used as the ground truth.

\subsection{J-PAS mock galaxy catalogues with FLAMINGO}

Since the J-PAS survey does not yet have enough spatial coverage to allow the study of cosmic voids,  we decided to use the Full-hydro Large-scale structure simulations with All-sky Mapping for the Interpretation of Next Generation Observations \citep[FLAMINGO,][]{2023MNRAS.526.4978S, 2023MNRAS.526.6103K, 2026arXiv260424324H} to mimic the expected J-PAS data. This simulation was selected for its large volume and hydrodynamical framework, which simultaneously allowed us to investigate large structures such as voids and the properties and evolution of individual galaxies residing within them.

 FLAMINGO is a suite of hydrodynamical simulations designed for cosmology and large-scale structure studies, building upon the BAHAMAS simulation \citep{2017MNRAS.465.2936M}. Stellar and AGN feedback models are calibrated to match the observed low-redshift galaxy stellar mass function (SMF) and cluster gas fractions, using Gaussian process emulators for efficient machine-learning-based optimisation. The two flagship runs cover volumes of 2.8 Gpc$^{3}$ and 1 Gpc$^{3}$ and baryonic particle masses of 1 $\times$ 10$^{9}$ $M_{\odot}$ and 1 $\times$ 10$^{8}$ $M_{\odot}$, respectively. The simulations use up to 3 $\times$10$^ {11}$ particles, one of the largest numbers of resolution elements for cosmological hydrodynamical simulation runs to z = 0 to date.
 The cosmological parameters used in the fiducial simulations are from the Dark Energy Survey year three \citep{PhysRevD.105.023520}, $\Omega_\Lambda = 0.694$, $\Omega_m = 0.306$, $\Omega_b = 0.0486$, $\sigma_8 = 0.807$, $n_s = 0.967$, and $h = 0.681$.

\begin{figure}
    \centering
    \includegraphics[]{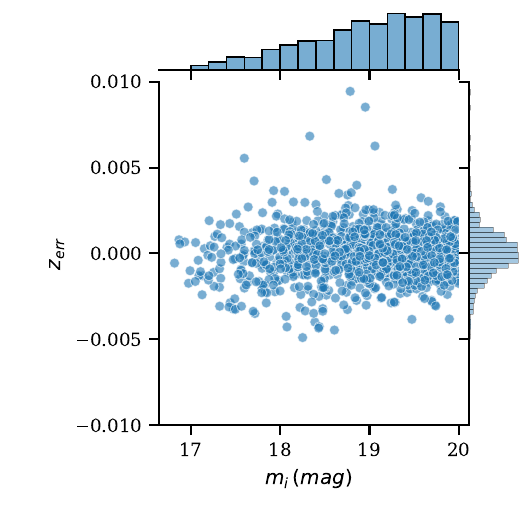}
    \caption{J-PAS photometric redshift error distribution provided by the TOPz pipeline (see text for details).}
    \label{z_err}
\end{figure}

To construct J-PAS mock catalogues, we first assess the survey's photometric redshift uncertainties, which are the primary limiting factor in void identification. For this purpose, we use the J-PAS Early Data Release EDR202411, which provides photometry in 54 narrow-band filters, two intermediate-band filters, and the $i$SDSS band for 26168 galaxies over a 17~deg$^{2}$ area. We define a J-PAS calibration sample by selecting galaxies in the redshift range $0.2 < z < 0.4$ that have an apparent magnitude limit in the i-band $m_{i} < 20$, with photometric redshifts estimated using the TOPz workflow. Furthermore, we restrict our sample to
galaxies with well-defined primary peaks in their redshift probability density functions. This criterion, along with the selected redshift range, is chosen to achieve a balance between minimising the median redshift error and maximising the total galaxy count. Given these restrictions, we note that the J-PAS calibration sample corresponds to a high-quality redshift J-PAS sample, and it might not be entirely representative of the total J-PAS galaxy population. In Fig.~\ref{z_err}, we show the relation between apparent magnitude and photometric redshift errors of the J-PAS galaxy sample, comprising 1140 galaxies after applying the aforementioned selection criteria. The median and 95th percentile of the absolute photometric redshift error distribution are 8.7 $\times 10^{-4}$ and 2.8 $\times 10^{-3}$.

To mimic these J-PAS observations, we use the high-resolution fiducial simulation (L1$\_$m8) of FLAMINGO that covers a volume of 1 Gpc$^{3}$ and has a baryonic particle mass of 1.34 $\times$ 10$^{8}$ $M_{\odot}$. This selection represents a necessary trade-off between achieving the large volume required for significant void statistics and the particle resolution required for realistic galaxy properties within a reasonable range of galaxy masses. We convert the luminosities of galaxies in FLAMINGO to apparent magnitudes, and we select a snapshot at $z_{\rm FLAMINGO}$ = 0.3 to match the magnitude distribution of the J-PAS galaxies.


 The galaxy properties (e.g., luminosities and star formation rates) are provided only for subhalos containing at least 100 star particles. We therefore restrict our sample to these systems, which have a minimum stellar mass completeness limit of $M_{\star} \geq 10^{10}\,M_{\odot}$. Consequently, our results characterise the population of relatively massive void galaxies, and we caution that the properties of the lower mass galaxies are not covered. We note that this limitation roughly corresponds to the J-PAS calibration sample's magnitude limit.

We model the expected redshift errors following the procedure in \cite{2025A&A...695A.174M}. The main steps are summarised below:

\begin{enumerate} 
    \item Obtain the photometric redshift errors $z_{\rm err}$ (Eq.~\ref{zerr}) of the J-PAS calibration sample, as provided by the TOPz pipeline.
    
    \item Add the J-PAS calibration sample redshift errors to the redshift of FLAMINGO snapshot, $z_{\rm FLAMINGO} = 0.3$, in order to obtain the J-PAS calibration sample redshifts, $z_{\rm pert}$:
    \begin{equation}
        (1 + z_{\rm FLAMINGO}) (1 + z_{\rm err}) = (1 + z_{\rm pert}).
    \end{equation}
    
    \item Convert the J-PAS calibration sample redshifts, $z_{\rm pert}$,  to corresponding comoving distances, $d_{\rm pert}$.
    
    \item Compute the distance errors, $d_{\rm err}$ , by subtracting the J-PAS calibration sample comoving distances from the fixed comoving distance at $z=0.3$, $d_{z=0.3}$:
    \begin{equation}
        d_{\rm err} = d_{z=0.3} - d_{\rm pert}.
    \end{equation}
    The resulting median and the 95th percentile of the absolute distance error distribution are 3.8~Mpc and 13.9~Mpc, respectively.
    
    \item Apply the inverse transform sampling technique to $d_{\rm err}$ and add the resulting offsets to the $z$-axis coordinates of FLAMINGO galaxies.  
\end{enumerate}

\begin{figure*}[t]
    \centering  
    \includegraphics[]{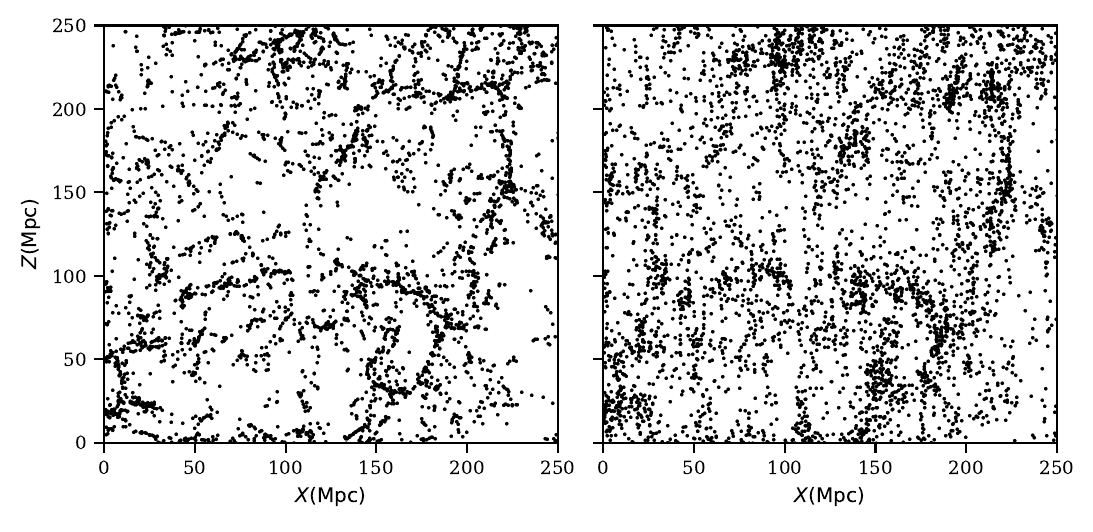}
    \caption{A 15~Mpc thick slice zoom-in of the 1~Gpc FLAMINGO box of the \textit{FBI} (left) and \textit{JP} (right) galaxy mocks. The effect of the photometric redshift errors on the galaxy distribution can be observed in the right panel.}
    \label{gal_slice}
\end{figure*}

Following this procedure, we construct two J-PAS galaxy mock samples: a FLAMINGO-based ideal (\textit{FBI}) sample, consisting of FLAMINGO galaxies with original, unaltered positions, and a FLAMINGO-based \textit{JP} sample, in which the galaxy coordinates are displaced along the $z$-axis to emulate the photometric redshift errors expected in the J-PAS survey. We do not model redshift-space distortions caused by peculiar velocities, as the resulting line-of-sight displacements are smaller than the modelled photometric redshift errors. While a complete treatment of observational systematics, such as survey geometry, selection functions, and angular masks, is essential for application to real data, we leave these aspects to a future work and focus here on a controlled analysis of the impact of photometric redshift errors.
In  Fig.~\ref{gal_slice}, we show plots of the \textit{FBI} and \textit{JP} galaxy mock samples in a 15~Mpc-thick slice. The smearing of the spatial galaxy distribution caused by redshift errors is visible in the right panel.

\section{Analysis and processing}
\label{sec:meth}

For the analysis of the void population in the maps produced by the
photo-$z$ survey, we use a sequence of processing steps.

Firstly, we translate the (discrete) galaxy distribution into a continuous number density field. To this end we invoke the
DTFE, which assures the representation
of the detailed geometric and multiscale spatial patterns of the cosmic web.

Next, we follow up with a two-tier strategy for the study of voids and void galaxies in photo-z surveys. In the first tier, we identify dynamically dominant (super)voids in the quasi-gravitational potential using the watershed formalism. Subsequently, we infer properties such as sizes, shapes and density
profiles of the extracted gravitationally dominant supervoids. Also, we
assess the individual recovery rate and characteristics of the recovered voids, comparing  their properties with the properties of the overall void population.

In the second tier, we define our samples of void galaxies and galaxies located in high-density regions on the basis of the density field. We also provide the number of void galaxies which are common to both, the \textit{FBI} and \textit{JP} samples.

\subsection{DTFE density field maps}


The simulation particle distribution is transformed into a continuous and volume-filling density field by means of DTFE~
\citep{2000A&A...363L..29S, 2009LNP...665..291V, 2011arXiv1105.0370C}.
For delineating underdense regions, it is necessary to process the discrete galaxy distribution into a continuous
density field that optimally reflects the underlying weblike cosmic matter distribution.

To this end, DTFE exploits the density and shape adaptive properties of Voronoi and Delaunay tessellations to retain the multiscale,
geometric and topological nature of the
mass distribution probed by the galaxy distribution. The Voronoi tessellation involves the division of space into mutually disjunct
polyhedra, each polyhedron consisting of the region of space closer to the defining point than any of the other points \citep{Voronoi1908, 2000stca.conf.....O}. The Delaunay tessellation is the dual of the Voronoi tessellation, with each Delaunay tetrahedron being defined as the set of four sample points whose circumscribing sphere does not contain any other of the sample points \citep{Del34,2000stca.conf.....O}. 

DTFE includes two key steps: (1) estimating the density value at the location of simulation particles or survey galaxies
and (2) subsequently interpolating these densities throughout the probed volume. In the first step, DTFE uses the density and shape adaptive
virtues of Voronoi tessellations. The volume of Voronoi cells is inversely proportional to the local number density of probes. To ensure mass conservation, the DTFE density estimate is defined as inversely proportional to the volume of the contiguous Voronoi cell (or
Voronoi star), which is the agglomerate of all Delaunay tetrahedra contiguous to the corresponding Voronoi cells, meaning the centre of each constituent Delaunay cell defines a vertex of the resulting Voronoi polyhedron. In the second step, DTFE interpolates the estimated density values at the sample locations over the entire sample volume by using the corresponding Delaunay tessellation as linear interpolation grid. In each Delaunay tetrahedron, the field gradient is assumed to be constant and determined, uniquely and completely, from the four field values of
the four defining sample points. It yields a volume-covering continuous density field. Finally, the value of the field at each grid
cell is computed as the volume-weighted average over the grid cell. The full DTFE procedure and implementation also includes the
option of interpolating the velocity field values, and yielding a volume-covering map of the volume-weighted divergence and shear
of the velocity flow field. 

The main advantage of DTFE is that it preserves the geometric features of voids, filaments, and clusters, while producing a continuous,
volume-weighted density field. Particularly important is that it outlines the presence and shape of voidlike regions and is also capable of producing a continuous flow field reconstruction throughout the sparsely sampled void regions. This has also been demonstrated in a range of DTFE applications to observational samples, in particular concerning the SDSS galaxy survey \citep{2010MNRAS.408..897J,2011MNRAS.416.2494P}. The 
study by \cite{2011MNRAS.416.2494P} entails a detailed quantitative study of the performance of DTFE in analysing the cosmic web in the SDSS galaxy survey, and in particular in recovering the void population. It combines this with an extensive study of extensions and higher order versions of
DTFE, notably involving Kriging and Natural Neighbor Interpolation \citep{sibson1981,Watson92}.

In the present study we use the public C++ implementation of DTFE \citep{2011arXiv1105.0370C}.
We run the DTFE on our two J-PAS galaxy mock samples with a grid spacing of 1 Mpc, assuming periodic boundary conditions. In the central column of Fig.~\ref{maps}, we show the logarithmic density field map for our two J-PAS mocks: \textit{FBI} (top) and \textit{JP} (bottom). Comparing these plots with the galaxy distributions in the left column, we observe that the high-intensity yellow features trace the overdense structures, while the dark blue regions delineate the underdensities.

\subsection{Gravitational potential watershed voids}

Photo-$z$ errors tend to strongly affect the obtained large scale structure observed by photometric surveys. The considerable photo-$z$ uncertainties
translate into a redshift map which is blurred accordingly along the redshift dimension, yielding a fuzzy image of the
underlying mass distribution (see Fig.~\ref{gal_slice}). In order to ameliorate the impact of the photo-$z$ distortions we choose
to focus only on the dynamically dominant voids in the cosmic mass distribution, the \textit{Superhubble Bubbles} \citep{1984MNRAS.206P...1I}
that are expanding along all directions and whose interiors are only mildly affected by the gravitational influence of exterior
structures - voids, walls, filaments and cluster nodes - that surround them \cite[for a recent discussion see][]{2026MNRAS.547ag193K}.
It involves the recognition that the large expanding voids that represent the expansion centres in the cosmic matter distribution should be well identifiable even in the case of major redshift uncertainties such as those encountered in photo-$z$ surveys.
It also means we will not include the majority of void regions, the usually smaller (sub)voids populating the outer regions of
these major expansion regions or the ones that are gravitationally dominated by their peers or surrounding filaments, walls and cluster nodes \citep[see][for a thorough discussion]{2026MNRAS.547ag193K}, which make them contract along one or more directions and may
even lead them to full collapse \citep{2004MNRAS.350..517S}. 
  

In the present study we therefore apply the watershed identification of voids in a quasi-gravitational potential field \citep{Mccarthy2024}. It adheres to the definition of superstructure attraction basins in the
gravitational potential \citep[e.g.][]{2014Natur.513...71T, 2020MNRAS.493.3513D,2023A&A...678A.176D,2025A&A...704A.151E}, such as Laniakea, by means of a watershed definition parallel
to the original watershed application to the density field \citep{2007MNRAS.380..551P}. For the gravitationally dominant voids we use
a similar strategy, by outlining the descending gravitational potential manifold around peaks in the gravitational potential
field.

Because the potential field is effectively a low-pass filtered version of the cosmic mass distribution, it only contains
the large gravitationally significant regions. Each of these regions encompasses the major share of small structures with
a limited gravitational influence. The redshift attenuation by the photo-$z$ uncertainties tends to be relatively
minor, which implies the identified void "basins" to be far less affected than the voids identified in the
density field. As described in the present study, we indeed find that these supervoids are quite well outlined in
photo-$z$ maps, i.e. their properties correspond well to the same voids in the underlying physical mass distribution.

\bigskip
\bigskip

\noindent\textbf{Gravitational potential field}\\

The gravitational potential, $\Phi$, can be computed using Poisson's equation:
\begin{align}\hspace{0.3cm}
    \nabla^2 \Phi = 4 \pi G \bar{\rho} \delta(\textbf{r})
    \label{poisson}
\end{align}
where $\bar{\rho}$ is the mean galaxy number density in the simulation box and $\delta (\textbf{r})$ is the density contrast, given by DTFE.

It is more convenient to express Eq.~\ref{poisson} in Fourier space such that Poisson's equation becomes
\begin{align}\hspace{0.3cm}
    \Phi(\textbf{k}) = -4\pi G \bar{\rho} \frac{\delta(\textbf{k})}{k^2}.
    \label{poisson_fourier}
\end{align}

To enhance the contrast of the potential field while accounting for the highly non-Gaussian nature of the evolved density field, thereby amplifying the relative significance of the low-density regions,
\citep[see eg.][]{1991MNRAS.248....1C, theses_fse8301,theses_fse33679,theses_fse36188}, we compute a
\textit{quasi-gravitational potential} by solving the Poisson equation for
the log-transformed density field $\ln(1 + \delta)$. The quasi-potential preserves the topological and dynamical
information of the potential field, and allows a balanced evaluation of both void regions and supercluster basins. We solve Eq.~\ref{poisson_fourier} on a grid with a spatial resolution of 1~Mpc, assuming periodic boundary conditions. 


From the potential point of view, gravitationally repulsive regions with $\Phi > 0$ correspond to major underdense
expansion and matter outflow regions. The potential basin $\Phi < 0$ regions typically concern regions into which matter is
attracted, accreted and accumulated. The low-pass nature of the gravitational potential implies it to be closer to its
primordial Gaussian nature, in which the $\Phi<0$ and $\Phi>0$ regions comprise comparable fractions of the
survey volume. This is quite different from the situation for the matter density field, in which voids occupy
up to $75\%$ of the cosmic volume \citep[see e.g.][]{2014MNRAS.441.2923C,Ganeshaiah_Veena_2019}. It is a direct reflection of the
\textit{void-in-cloud} phenomenon, in which a large fraction of (small) voids find themselves embedded in
gravitationally contracting and even collapsing regions, often corresponding to larger scale overdensities
\citep{2004MNRAS.350..517S}. In other words, by restricting our attention to the gravitationally repulsive $\Phi>0$ regions, we
concentrate on the fraction of voids that dominate their immediate surroundings, and which have formed and hierarchically
evolved through the \textit{void-in-void} process. 

\begin{figure*}[t]
    \centering
    \begin{minipage}[t]{\textwidth}
        \centering
        \includegraphics[]{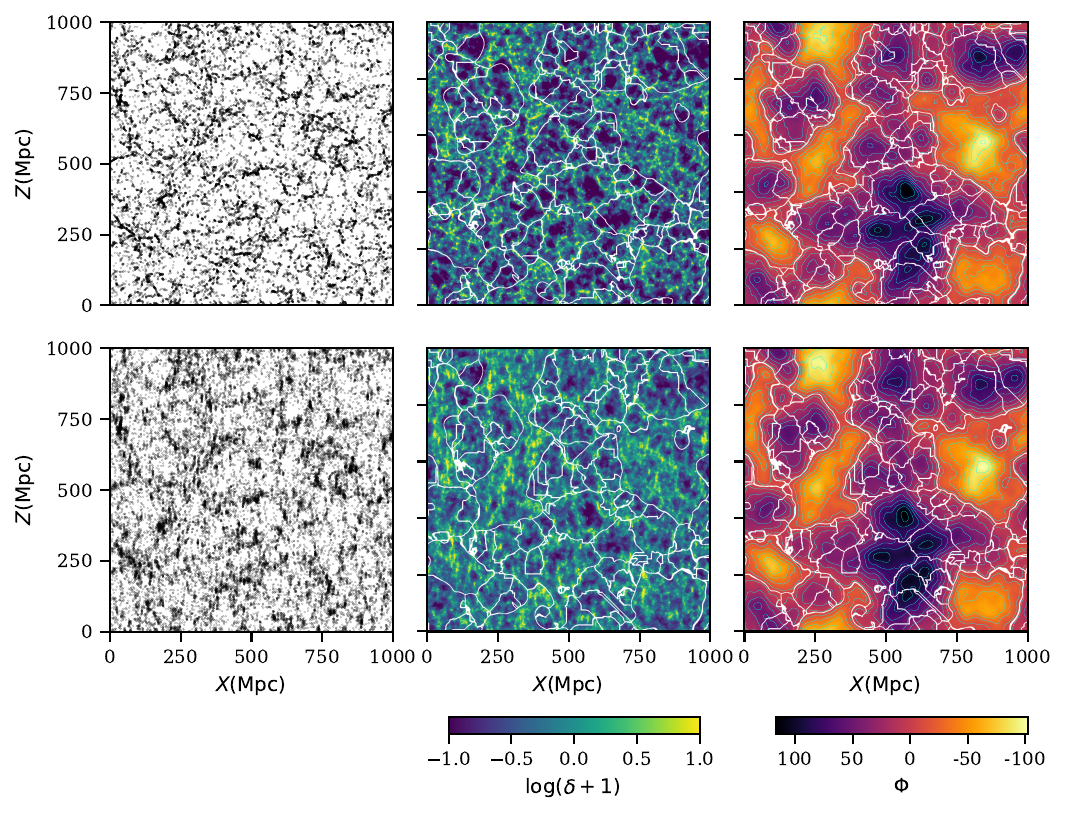}
        \caption{\textit{FBI} (top row) and \textit{JP} (bottom row) galaxy mocks of J-PAS shown in a 5~Mpc-thick slice (left column), together with the corresponding density fields (middle column) and the quasi-potential fields (right columns) displayed in a 1~Mpc-thick slice in the $X$–$Z$ projection. White lines indicate the watershed void boundaries. Bright regions in the quasi-potential field map correspond to regions of $ \Phi < 0$ which have been excluded from the void identification. Equal quasi-potential value contours are shown with dotted lines.}
    \label{maps}
    \end{minipage}
    \hfill
\end{figure*}

For visual appraisal, Fig.~\ref{maps} shows the density fields (central column) and the quasi-potential fields (right column) for the
\textit{FBI} galaxy sample and the photo-$z$ distorted \textit{JP} sample. The bright-coloured regions correspond to $\Phi<0$ (collapsing regions) while the darker ones to $\Phi>0$ (expanding regions). The figure shows a clear correspondence
between the structures in the density field and their lowpass filtered counterparts in the gravitational
potential field. It reflects the effect of the lowpass $1/k^{2}$ factor in Eq.~\ref{poisson_fourier}, which
suppresses small-scale fluctuations while preserving only the large-scale modes.

\bigskip
\bigskip
\noindent\textbf{Gravitational potential watershed voids}\\

\noindent Following the computation of the quasi-potential field, we apply the watershed formalism to detect the
void regions following the detailed outline of the technique in the original study introducing the
watershed formalism for void detection \citep{2007MNRAS.380..551P}. The modification of the watershed
formalism towards the identification in the quasi-potential fields is extensively described in \cite{Mccarthy2024}. 

The watershed identifies local minima in the negative of the quasi-potential (negative quasi-potential, for short). We use the negative quasi-potential rather than the regular quasi-potential to keep the analogy between void regions and basins
in the context of the watershed procedure: voids correspond to basins in the landscape of the negative quasi-gravitational potential. The
minima in the negative quasi-potential are the seeds, or expansion centres \citep{1984MNRAS.206P...1I}, of the individual voids. Subsequently, void
basins are outlined by following the integral lines emanating from the corresponding minima in the field. The integral lines
represent the local gradient along the field. In typical watershed void finders, these are the density field gradient
lines, while in the potential field the integral lines are in fact the flowlines in the mass distribution\footnote{This was in
fact the rationale for the identification of the Laniakea supercluster complex, and its peers, from the flowfield
inferred from the Cosmicflows peculiar velocity survey \citep{2014Natur.513...71T, 2020MNRAS.493.3513D,2023A&A...678A.176D}.}. A void basin is
then identified with the ascending manifold surrounding the minimum \footnote{These are mathematical concepts 
defined and described by Morse theory, see \cite{morse1934,milnor1963}.}, the set of points belonging to the integral lines
whose origin is that of the corresponding minimum in the negative quasi-potential field. The boundaries between two adjacent void basins
are the integral lines marked by the presence of a saddle point and stretching between two adjacent maxima. These are 
so-called \textit{separatrices}.

In practice, the integral lines and separatrices are directly determined from the quasi-potential field values at the grid points of
the grid on which the field has been evaluated.  The algorithm identifies local minima in the negative quasi-potential grid, acting as the
seeds and expansion centres of individual voids. In each iteration, neighbouring cells are examined: if a neighbour has a
higher negative quasi-potential value than the local minimum, it is assigned the same label as that minimum, effectively allowing the
basin to “grow” outward. This process is repeated iteratively, with the negative quasi-potential flowing downhill toward the nearest
minimum, until convergence is reached and all cells are assigned to distinct basins separated by watershed boundaries
(see Fig. 1 from \cite{2007MNRAS.380..551P} for a good illustration of the watershed principle).

We seek to minimise the effect of photometric redshift errors by restricting the void population to the dynamically
dominant expanding regions in the cosmic web, and to prevent including the (smaller) voids susceptible to photo-$z$
redshift errors as they are embedded within overdense regions or dynamically dominated by their environment
\citep{2025A&A...695A.174M}. For that, we impose a threshold for the watershed growth process such that we exclude regions for which the quasi-potential corresponds to -$\Phi~\geq~0$, which are
expected to be associated with collapsing or contracting features. Imposing such a threshold results in
a void volume occupation of $\sim$ 55$\%$ in the \textit{FBI} sample and  $\sim$ 56$\%$ in the\textit{ JP} sample. 

By virtue of the method, the watershed algorithm partitions the entire volume into distinct basins -- regions of locally low negative quasi-potential -- separated by ridges corresponding to higher negative quasi-potential (and, correspondingly, higher-density) boundaries. Because this does not involve \textit{a priori} geometric assumptions about the size or shape of the voids, the resulting
basins have been grown naturally and yield a diverse population of voids with a range of sizes and anisotropic shapes.

 In the middle and right columns of Fig.~\ref{maps}, we display cross-sections of voids from the \textit{FBI} (top) and
 \textit{JP} (bottom) samples. The watershed boundaries are indicated by white contours, overlaid on the density and quasi-potential fields within a slice of thickness 5 Mpc. In the quasi-potential maps, darker regions correspond to higher values ($\Phi > 0$), while brighter regions indicate lower values ($\Phi < 0$).

We note that this paper concerns an application of the quasi-potential framework to J-PAS galaxy mocks, rather than a methodological validation paper of the quasi-potential method. A quantitative calibration and sensitivity suite addressing optimal parameters (including grid resolution, log-transforms, and quasi-potential thresholds) is the core subject of a dedicated companion methods paper currently in preparation (McCarthy et al., in prep.).    
 
 
\subsection{Void properties}
\label{sec:void_prop_meth}
We evaluate a range of structural characteristics for the voids identified by means of the quasi-potential watershed procedure, and assess their statistical distribution. These include the void centre, equivalent radius, their shape and the void density profiles. 

Important for our confidence in the obtained results, is the evaluation of the success with which we are able to recover individual voids in the \textit{FBI} mock from the \textit{JP} mock using the quasi-potential watershed procedure. Hence, we also evaluate the individual \textit{void recovery}.

\bigskip
\noindent \textbf{Void centre} \\
\label{sec:sizes_sect}

The definition of a void centre remains somewhat ambiguous, as it can vary depending on the context of the study  \citep[e.g.][]{2015MNRAS.454.2228N, 2021MNRAS.505.1223P}. Some possible choices include the geometric centre, the location of minimum density, or the point of maximum potential from which matter flows outward. These locations do not necessarily coincide and are often offset from one another by several megaparsecs within the same void.

We define the \textit{geometric centre} of a void by computing the independent median of the $x$, $y$, and $z$ coordinates of all
void grid cells. We also find the location of the \textit{maximum quasi-potential point} within each void. The different void centre definitions, either geometric centre or maximum quasi-potential points, are used in different
situations. For example, the geometric centres are used when computing the void shapes, while
the maximum quasi-potential points are used for computing the spherical density profiles and the identification of
void galaxies in the interior of voids.

\bigskip
\noindent \textbf{Void size} \\
\label{sec:sizes_sect}

\noindent We compute the void sizes in two complementary ways. First, we adopt the conventional approach by using the equivalent radius, which is defined as the radius of a sphere that has the same volume as the volume of a void:
\begin{align}\hspace{0.3cm}
    R_{\rm eq} = \left(\frac{3}{4 \pi} V \right)^{\frac{1}{3}},
    \label{eqrad}
\end{align}
where $V$ is the volume of a void, obtained as the sum of all grid cells belonging to that void. However, since the watershed algorithm naturally produces anisotropic basins, approximating such irregular shapes with spheres represents only a first approximation \citep{2016MNRAS.457.2540C}.

In order to obtain a better void size estimator, we compute the distance from the void boundary, following the recipe of \cite{2016MNRAS.457.2540C}. The minimum distance, $\mathcal{D}$, from the void boundary to a point with a coordinate $\textbf{x}$ is given by:
\begin{align}
\mathcal{D} =
\begin{cases}
+\min_i \lVert \textbf{x} - \textbf{y}_i \rVert & \text{for } \textbf{x} \text{ outside the void}, \\[6pt]
-\min_i \lVert \textbf{x} - \textbf{y}_i \rVert & \text{for } \textbf{x} \text{ inside the void},
\end{cases}
\label{disttrans}
\end{align}
where {$\textbf{y}_i$} represents the set of points that defines the void boundary and || denotes the magnitude of a vector. The boundary distance is negative inside the void and positive outside. Furthermore, for each void in the two galaxy mocks, we define a boundary distance field (the distance transform), $\bm{\mathcal{D}}$, as a vector of magnitude $\mathcal{D}$, given by:
\begin{align}
\bm{\mathcal{D}} = \mathcal{D} \frac{\mathbf{x} - \mathbf{y}_j}{\lVert \mathbf{x} - \mathbf{y}_j \rVert},
\label{dist_field}
\end{align}
where $j$ denotes the index of a point on the void boundary closest to \textbf{x}. For each void, there is a point that takes a minimum value, $\mathcal{D}_{min}$, and is located farthest from the boundary. However, since referring to this point as minimum boundary distance might generate confusion, we define the maximum boundary distance, $\mathcal{D}_{max}$, as the absolute value of the minimum distance found within the void ($\mathcal{D}_{max} = |\mathcal{D}_{min}|$). Since the voids produced by the watershed void finder are non-spherical, it follows that $\mathcal{D}_{max}$  $\leq R_{\rm eq}$.\\

\begin{figure}[]
    \includegraphics[]{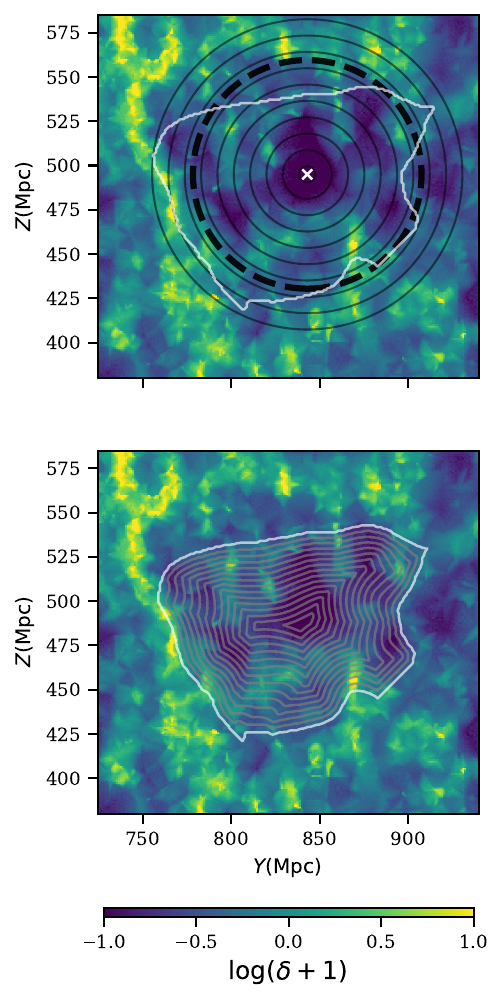}
    \centering  
    \caption{Cross-section of a watershed void identified in the quasi-potential field of the \textit{JP} mock overlayed on the density field. The white contour represents the watershed void boundary. Top panel: the black concentric circles, centred on the maximum quasi-potential point (cross), illustrate the sampling spherical shells used for the computation of the spherical density profile. The thick, dashed circle has a radius equal to the equivalent radius of the shown void.
    Bottom panel: The grey contours represent lines of equal distance from the boundary, used to compute the boundary density profile.}
    \label{void_bd}
\end{figure}

\bigskip
\noindent \textbf{Void shape} \\

\noindent We estimate the shapes of voids with ellipsoids by computing the shape-tensor $S_{ij}$ \citep{2008MNRAS.387..128P}. For each void, the tensor was computed by summing over the $N$ void volume elements $k$:
\begin{align}
S_{ij} &= - \sum_k x_{ki} x_{kj} && \text{(off-diagonal)}, \\
S_{ii} &= \sum_k \left( \textbf{x$_k$}^2 - x_{ki}^2 \right) && \text{(diagonal)} \,,
\end{align}
where $\textbf{x}_k$ is the position of the $k$-th volume element within the void, relative to the geometric centre.

The eigenvalues of the shape ellipsoid, $s_i$ with $s_1 > s_2 > s_3$, were used to obtain the lengths of the semi-axes $a_i$:
\begin{align}
        a_1^2 &= \frac{5}{2N} (s_2 + s_3 - s_1) , \\
        a_2^2 &= \frac{5}{2N} (s_1 + s_3 - s_2) , \\
        a_3^2 &= \frac{5}{2N} (s_1 + s_2 - s_3) ,
\end{align}
where $N$ is the number of volume elements and with $a_1 \geq a_2 \geq a_3$. The shape of voids are quantified via two axis ratios $\eta_{21} = a_2 / a_1$ and $\eta_{32} = a_3/a_2$ and the ellipticity, $\epsilon = 1 - \eta_{31}$, with $\eta_{31} = a_3/a_1$. We distinguish between oblate voids, those with $\eta_{21} > \eta_{32}$, and prolate ones, with $\eta_{21} < \eta_{32}$. A perfectly spherical void has $\eta_{21} = \eta_{32} = 1$. \\

\bigskip
\noindent \textbf{Void density profile} \\

\noindent In order to probe the number density within voids, we compute two types of void density profiles: the spherical density profile (inside-out) and the boundary density profiles (outside-in). We derive the spherical density profile of each void by averaging the density field values within concentric shells centred on the point of maximum quasi-potential. The individual profiles are subsequently scaled to each void’s equivalent radius. The stacked density profile is obtained by computing the average of the individual profiles within each radial bin. The boundary density profile follows the method of \cite{2016MNRAS.457.2540C} and uses the distance transform defined in the previous section (Eq.~\ref{disttrans}). It is obtained in a similar way to the spherical profile, except that the averaging of the density field values takes place in boundary-shaped shells within the void.

In the top panel of Fig.~\ref{void_bd}, we illustrate, as an example, the concentric spherical shells used to compute the spherical density profile for a void in the \textit{JP} mock. The thick, dashed circle has a radius equal to the equivalent radius of the void. When computing the spherical density profile, one can notice how the spherical shells may end up sampling the density field outside of the void. The bottom panel shows the boundary distance field for the same void. By taking into account the shape of the void, there is no danger of sampling the density field beyond its outer parts. That being said, the location situated at the maximum boundary distance does not have to coincide with the lowest density region within the void. This might cause higher values than expected for some voids in their boundary density profiles' interiors (see Fig.~\ref{bdprof} in our results section).\\

\noindent \textbf{Void recovery} \\

One of our goals is to identify corresponding voids between the \textit{FBI} and \textit{JP} samples. We refer to this operation as void recovery. To recover voids, we utilise their volumes, previously obtained using the watershed void finder. We consider a void to be recovered if the intersection over union (IoU) is larger than 0.5, where the IoU is defined as
\begin{align}
    \text{IoU} = \frac{V_{overlap}}{V_{FBI} + V_{JP} - V_{overlap}},
\end{align}
with $V_{overlap}$ representing the volume overlap between a void in the \textit{FBI} sample and a void in the \textit{JP} sample while $V_{FBI}$ and  $V_{JP}$ being the void volumes in the respective samples. IoU ranges from zero (no overlap) to one (perfect overlap).

\subsection{Void galaxy selection and properties}
\label{sec:voidgals}

While the identified supervoids delineate the gravitational influence regions, for the study of void galaxies, we
follow this up with a two-tier strategy, given that this involves how the low-density environment influences the formation, evolution and fate of galaxies.

\begin{figure}[]
    \includegraphics[]{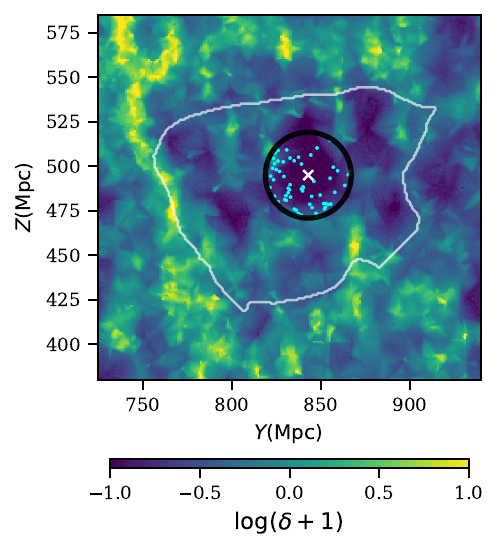}
    \centering  
    \caption{Cross-section of a watershed void identified in the quasi-potential field of the \textit{JP} mock overlayed on the density field.  The white contour represents the watershed void boundary. The black circle has a radius equal to the radius of a sphere, centred on the maximum quasi-potential point (cross), which contains 50\% of the void's average density. Cyan points within the circle are void core galaxies, as defined in Sect.~\ref{sec:voidgals}.}
    \label{void_core_gal}
\end{figure}

Hence, we supplement the quasi-potential watershed formalism with that of selecting resident galaxies
in spherical subregions in and around the interior centre of the identified quasi-potential voids. We define
\textit{void core galaxies} based on two criteria: (1) they must reside within a sphere centred on the maximum quasi-potential point that encloses 50\% of the specific void's mean density (concretely, the average density of DTFE cells comprising the volume of the void), and (2) they must satisfy a local density threshold of $\log_{10}(1 + \delta) < -0.3$, where $\delta$ is the density contrast given by the DTFE. In other words, the density value in the DTFE cells in which galaxies reside should not exceed 50\% of the global average density. This dual criterion isolates galaxies in the most rarefied environments while maintaining a sample size sufficient for statistical analysis. This selection has been applied independently to both the \textit{FBI} and \textit{JP} samples. In Fig.~\ref{void_core_gal}, we show an example of a void identified in the quasi-potential field of the \textit{JP} mock with its void core galaxies. The void and its galaxies are overlayed on top of the density field demonstrating that this selection guarantees that the void galaxies populate sparse regions of the void. 

To investigate the influence of environmental density on galaxy properties, we investigate whether for the identified void core galaxies
the familiar trends in galaxy properties, such as stellar mass, colour, and star formation activity (SFR and sSFR) can be observed
despite the photo-$z$ redshift errors. To this end, we compare the obtained galaxy properties with those in a comparison sample,
consisting of galaxies in higher-density regions with $\log_{10}(1 + \delta) > 0.5$. This high density threshold is intended
to assure that the selected comparison sample galaxies are not present within underdense regions and hence allow a fair comparison
between environments. The results of the void galaxy analysis are presented in Sect.~\ref{sec: void galaxies}.

\section{Voids in J-PAS mocks}
\label{sec:void_results}

 The primary goal of our work is to identify robust, dynamically relevant voids in J-PAS galaxy mocks by mitigating photometric redshift errors via a quasi-gravitational potential. We begin by characterising and comparing the general properties of voids -- specifically their abundance, sizes (Fig.~\ref{sizes}), density profiles (Figs.~\ref{sphprof} and \ref{bdprof}), and shapes (Fig.~\ref{shapes}), in both, the {\it FBI} and {\it JP} mocks. We evaluate the recovery of individual voids and briefly discuss how these voids deviate from the general population.

\label{sec:voidprop}

\subsection{Void abundance}
Using the watershed void finder algorithm, we identify voids in the quasi-potential fields of the \textit{FBI} and \textit{JP} mock catalogues. We find 1074 voids in the \textit{FBI} sample and 817 voids in the \textit{JP} sample.

In the middle and right columns of Fig.~\ref{maps}, we show cross-sections of voids delimited by their watershed boundaries, presented with solid white lines, in the density and the quasi-potential fields of the \textit{FBI} (top) and \textit{JP} (bottom) samples in a 5 Mpc/h thick-slice. Bright, orange regions in the right column correspond to regions of low quasi-potential, which we have excluded in the void identification via our quasi-potential threshold. We also show equal quasi-potential value contours to guide the reader towards the extremal locations. In both galaxy samples, voids present a variety of shapes and sizes, a consequence of the watershed nature to retain the topology of the field. A good correspondence between the voids in the \textit{FBI} and \textit{JP} samples can be observed. It is also worth noting that, since voids are identified in a quasi-potential field, rather than directly in the density field (as it is commonly done), the resulting watershed basins tend to be larger. A visual inspection of the density field within voids further reveals that several smaller underdense regions are nested within these broader basins. These may correspond to the void-in-void basins as predicted by the void hierarchy \citep{2004MNRAS.350..517S}.

\subsection{Void sizes}
\label{sect:void_size}

\begin{figure}[]
    \includegraphics[]{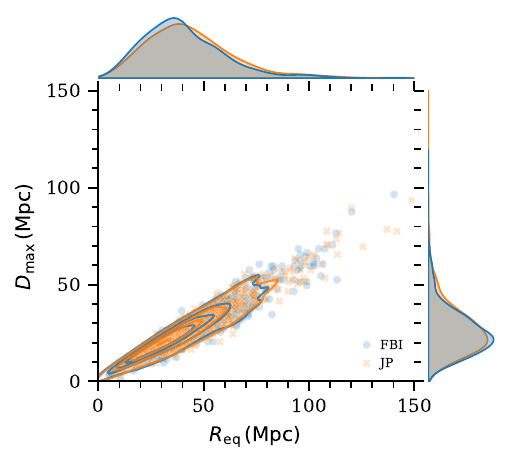}
    \caption{Void size distributions of the equivalent radius and the maximum boundary distance for the full void samples in the \textit{FBI} and \textit{JP} galaxy samples. Outermost contour lines enclose 95$\%$ of the data.}
    \label{sizes}
\end{figure}

In Fig.~\ref{sizes}, we show the distributions of the equivalent radius (Eq.~\ref{eqrad}) and the maximum boundary distance (Eq.~\ref{disttrans}) for voids in the two galaxy mocks. As expected, there is a strong correlation between the two parameters since both are size estimators. We find a wide range of void sizes, spanning 5-200~Mpc in equivalent radius and 3-150~Mpc in maximum boundary distance. For voids in the \textit{FBI} sample, the median values are 37~Mpc and 23~Mpc for the equivalent radius and maximum boundary distance, respectively, with slightly higher values of 40~Mpc and 24~Mpc in the \textit{JP} sample. We perform two separate two-sample Kolmogorov-Smirnov (KS) tests to determine whether the \textit{FBI} and \textit{JP} samples share the same underlying distribution for a given size estimator (equivalent radius and maximum boundary distance, individually). For the maximum boundary distance, we obtain a KS statistic of $D_n = 0.0603$ and a $p$-value of $p = 0.0651$. Since $p > 0.05$, we fail to reject the null hypothesis. While the obtained $p$-value does not strictly prove equality of the distributions, this result indicates no statistically significant difference between the maximum boundary distance of the two samples at the 5$\%$ significance level, allowing us to proceed under the assumption that they share a similar parent population. Conversely, for the equivalent radius, we obtain $D_n = 0.0889$ and $p = 0.0012$. Because this $p$-value is well below the significance threshold, we reject the null hypothesis and conclude that the two equivalent radius distributions differ. As noted in Sect.~\ref{sec:void_prop_meth}, the equivalent radius serves only as a first-order approximation of watershed void sizes and fails to capture their complex geometries. Thus, relying on the more robust maximum boundary distance estimator, we find good correspondence between the void size distributions in the two galaxy mocks. As noted in Sect.~\ref{sec:sizes_sect}, the equivalent radius typically exceeds the maximum boundary distance; an overestimation arising from the assumption that voids are spherical in shape. Concretely, we find that our void samples have a ratio between the average maximum boundary distance and the equivalent radius of $\mathcal{D}_{max}$ / $R_{\rm eq} \sim 0.64 \pm 0.09$ in the \textit{FBI} sample and $\mathcal{D}_{max}$ / $R_{\rm eq} \sim 0.63 \pm 0.09$ in the \textit{JP} sample, respectively. This result is in good agreement with the findings of \cite{2016MNRAS.457.2540C}.

The larger void sizes in the \textit{JP} sample suggest that positional uncertainties, combined with the smoothing of the quasi-potential, effectively inflate perceived void sizes. As tracers at the void boundaries scatter, the high-density ridges are diffused and subsequently smoothed; this suppresses the detection of small-scale voids and shifts the overall distribution in the \textit{JP} sample toward larger values. Despite these effects, the void size distributions remain largely consistent, confirming that the quasi-potential provides a robust framework for void identification that is relatively insensitive to photometric redshift errors.

The identification of voids and hence void properties, including their sizes, strongly depends on the tracer distribution, the smoothing scale \cite[e.g.][]{2015MNRAS.454..889N, 2017MNRAS.470.4434P, 2024MNRAS.529.4325B}, and void identification algorithm \citep{2008MNRAS.387..933C}. Moreover, given that our identification of voids takes place within a quasi-potential field rather than a density field, the average size of voids tends to be larger than what previous studies have found in simulations \cite[e.g.][]{2017MNRAS.470.4434P,2020MNRAS.493..899H, 2024ApJ...962...58C}. This can be attributed to two main factors. First, the quasi-potential effectively smooths small-scale structures that are otherwise prominent in the density field (most notably the sub-voids, hinting at the void hierarchy, that can be observed within the larger watershed regions in Fig.~\ref{maps}). Second, our quasi-potential threshold preferentially selects large, expanding voids, further shifting the size distribution toward higher values. Void sizes also depend on the cosmological model of choice. For example, \cite{2019JCAP...12..040V}  showed that voids identified in the dark matter density field are sensitive to different evolving dark energy equations of state, with the void abundance showing variations with respect to $\Lambda$CDM, the differences being strongest for smaller void sizes (specifically, for voids with an equivalent radius less than $R_{\rm eq} \lessapprox 25$~Mpc~$h^{-1}$. Given our voids are larger, on average, we do not expect different models to impact their sizes.




\subsection{Void density profiles}
Since we have identified basins in a quasi-potential field rather than a density field, we need to investigate whether our voids are truly underdense. We quantify the matter density of our voids by computing both spherical and boundary density profiles. 

\begin{figure*}[p] 
    \centering  
    
    \begin{minipage}[t]{\textwidth}
        \centering  
        \includegraphics[width=0.85\textwidth]{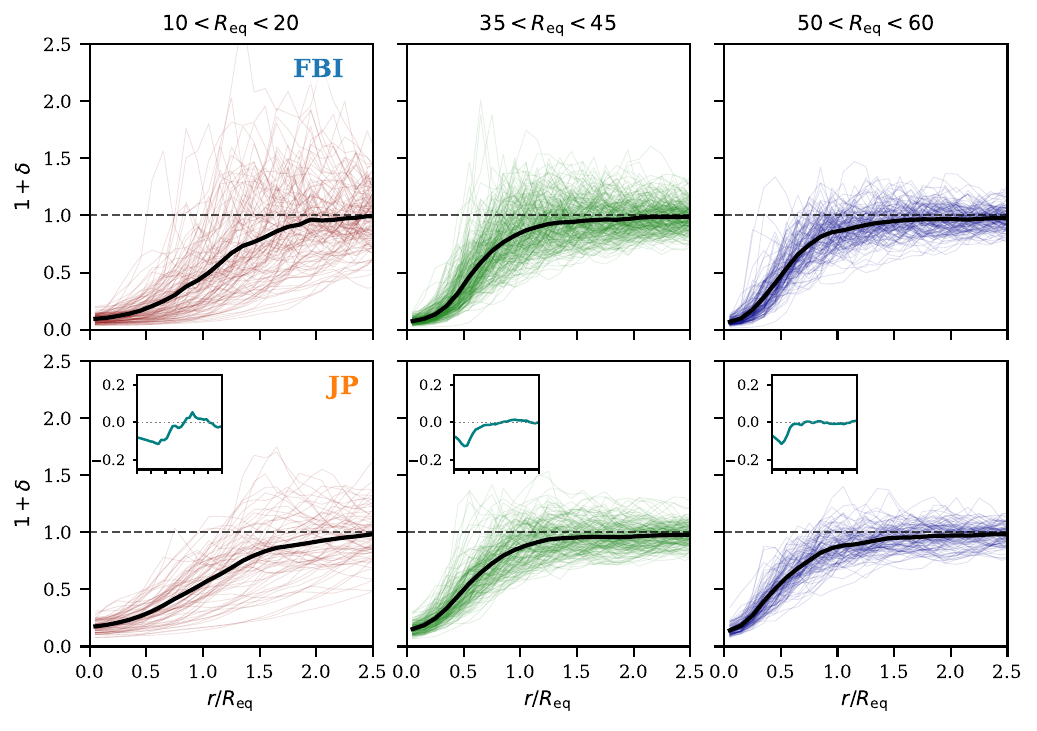}
        \caption{Individual (coloured lines) and stacked (black lines) spherical density profiles for \textit{FBI} (top) and \textit{JP} (bottom) sample voids, in three different equivalent radius intervals, in units of Mpc. Individual void profiles have been normalized to the equivalent radius of the respective void. The horizontal dashed line marks the average density, $1 + \delta$, in the simulation. Inset plots in the bottom row show the difference between stacked profiles for the \textit{FBI} and \textit{JP} sample voids.}
        \label{sphprof}
    \end{minipage}

    \vfill 

    \begin{minipage}[b]{\textwidth}
        \centering  
        \includegraphics[width=0.85\textwidth]{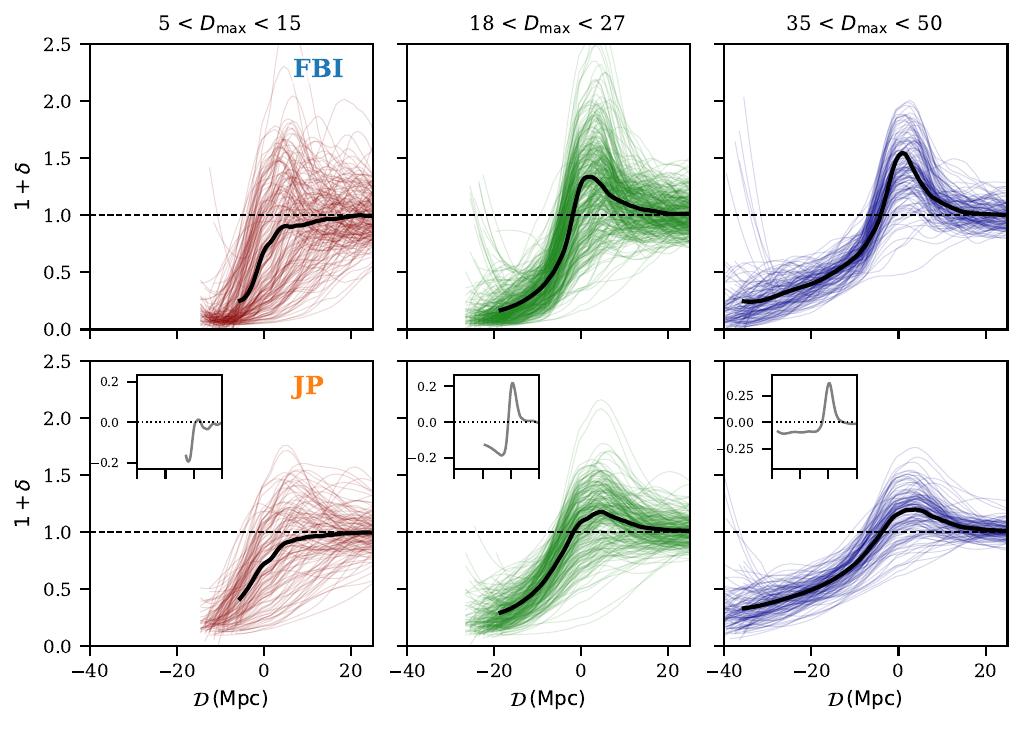}
        \caption{Individual (coloured lines) and stacked (black lines) void boundary density profiles are presented for the \textit{FBI} (top) and \textit{JP} (bottom) samples across three maximum boundary-distance bins. Stacked profiles are obtained by averaging the individual profiles to the lower bound of their respective boundary-distance interval.}
        \label{bdprof}
    \end{minipage}
\end{figure*}

Figure~\ref{sphprof} presents individual and stacked spherical density profiles for three different equivalent-radius bins in order to investigate how the matter content varies with void size. We show the profiles separately for the \textit{FBI} (top) and \textit{JP} (bottom) sample voids. \citet{2013MNRAS.434.1435C} distinguished voids arising from different evolutionary processes based on their density profiles. The authors differentiated between R-type voids, whose density profiles rise smoothly and monotonically from the centre toward the cosmic mean, and S-type voids, which exhibit an overdense shell. Dynamically, $S$-type profiles are characteristic of \textit{void-in-cloud} regimes, whereas $R$-type profiles are typical of \textit{void-in-void} configurations, though there are a range of other factors that
may strongly influence this classification. In our case, across all considered size ranges, the stacked density profiles indicate that both the \textit{FBI} and \textit{JP} voids identified within the quasi-potential framework belong to the $R$-type category. These structures remain distinctly underdense in their interiors, showing a gradual density increase that approaches the cosmic average only at a normalised radius of $r/R_{\rm eq} > 2$. This behaviour is expected, as our quasi-potential threshold inherently filters out \textit{void-in-cloud} systems, leaving a sample composed of deep, expanding underdense basins that directly neighbour one another.

The individual profiles present multiple overdense peaks ($1+\delta > 1$) at different scales, indicating the complex hierarchical sub-structure that resides within the quasi-potential field voids.  This is especially visible in the case of smaller voids (\(10 < R_{\mathrm{eq}} < 20\,\mathrm{Mpc}\)) which might neighbour a low quasi-potential (high-density) region. Another reason for the presence of the various peaks is the choice of void centre. The maximum quasi-potential within a void, which was chosen as the centre of the sampling spheres in computing the density profiles, does not necessarily coincide with the geometric centre. As such, at large radii, one can end up probing matter content that lies within neighbouring structures. The choice of void centre reinforces the point that applying spherical probes to a watershed void, whose geometry departs from sphericity, is inherently limiting.


Comparing the profiles of the \textit{JP} sample voids with those of the \textit{FBI} sample, one can notice a lower number of overdense peaks. The lower abundance of overdense peaks results from photometric redshift errors, which displace galaxies from high-density regions, thereby reducing the overall density contrast. This effect can also be observed as an increase in density in the interior of the \textit{JP} sample voids, visible especially in the inset panels of the bottom plots, where we show the difference between the stacked profiles of voids in the \textit{FBI} and \textit{JP} samples.

To better trace the matter content within the complex shapes of watershed voids, we additionally computed the boundary density profiles -- the density profiles moving in from the void boundary and following its shape (Fig.~\ref{void_bd}). In Fig.~\ref{bdprof}, we present both the individual and stacked boundary density profiles of voids in the \textit{FBI} and \textit{JP} samples, shown for three ranges of the void size. We notice even more clearly how the photometric redshift errors erase high-density structures, leading to voids with rather low-density boundaries and contamination of their inner regions, which manifests as a density increase observable in the \textit{JP} sample void profiles.

We compare these profiles to the corresponding spherical density profiles in order to highlight their similarities and differences. In both cases, the void interiors are typically underdense as shown by the rather flat density profiles, reaching typical values of $1+\delta \simeq 0.2$ for the boundary profiles and $1+\delta < 0.1$ for the spherical profiles. In some cases, we observe high-density values within void interiors in the boundary density profiles. This occurs because sampling deeper into a void using boundary shells does not guarantee that the smallest shells will capture the lowest density regions.

Despite the qualitative agreement in the void interiors, the two types of profiles exhibit clear differences at larger distances. In particular, the boundary density profiles of medium and large voids display a pronounced high-density ridge near the void boundary - a natural result indicative of the watershed ridge. Such a feature is absent in the spherical profiles, which instead show a more gradual increase in density towards the density mean. In contrast, small voids lack a prominent overdense boundary in their stacked profiles. This likely results from our quasi-potential threshold, which prematurely truncates the watershed growth before a typical ridge can develop. Such truncation occurs because these small voids are frequently adjacent to regions of low quasi-potential.



\subsection{Void shapes}

\begin{figure*}[t]
    \centering  
    \begin{minipage}[t]{\textwidth}
    \centering  
    \includegraphics[]{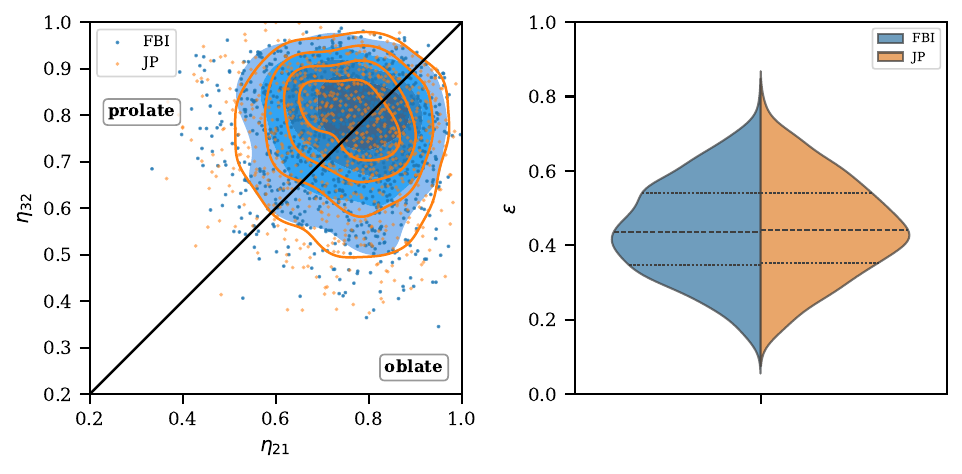}
    \caption{Comparison of shape parameters for voids in the \textit{FBI} (blue) and \textit{JP} (orange) mocks. The left panel shows the distribution of the fitted ellipsoid axis ratios. Voids lying on the left side of the diagonal ($\eta_{32} > \eta_{21}$) are prolate, while voids lying on the right side are oblate ($\eta_{32} < \eta_{21}$). The right panel displays the ellipticity distributions. Horizontal dashed lines indicate the 25th, 50th (median), and 75th percentiles for each distribution.}
    \label{shapes}
    \end{minipage}
    \hfill
\end{figure*}


Watershed voids can deviate substantially from perfect spheres, exhibiting a wide variety of shapes (see Fig.~\ref{maps}). In Fig.~\ref{shapes} we show the axis ratios (left panel) and the ellipticity (right panel) distributions of the fitted ellipsoids for voids in the \textit{FBI} and \textit{JP} samples as defined in Sect.~\ref{sec:void_prop_meth}. Indeed, we find that voids in both galaxy mocks are non-spherical, as evidenced by an average ellipticity $\epsilon \sim 0.44$. We also find median values of the axis ratios, $\eta_{21} = 0.75$ and $\eta_{32} = 0.77$ for the \textit{FBI} sample, and nearly identical results for the \textit{JP} sample ($\eta_{21} = 0.75, \eta_{32} = 0.76$), indicating consistent shapes across both mocks. We perform a two-sample KS test for the ellipticity and obtain a KS statistic  $D_n$ = 0.0365 and a $p$-value of p =  0.5519. Since $p > 0.05$, we fail to reject the null hypothesis, suggesting that there is no statistically significant difference between the ellipticity of the two samples at the 5$\%$ significance level.


We examined the relationship between void volume and ellipticity using the Spearman correlation coefficient ($r_s$). The resulting values ($r_s = 0.024$, $p = 0.430$ for the \textit{FBI} sample and $r_s = -0.008$, $p = 0.797$ for the \textit{JP} sample) indicate no statistically significant correlation. This might suggest that void volume and ellipticity are not associated in these samples.


\citet{2006MNRAS.367.1629S} investigated the shapes of voids in a $\Lambda$CDM simulation using fitted ellipsoids, finding average values of $\eta_{21} \sim 0.65$ and $\epsilon \sim 0.55$. Similarly, \citet{2008MNRAS.387..128P} examined the shapes and alignments of watershed voids, reporting ellipticity values with $\epsilon \sim 0.51$. The comparatively more spherical voids in our sample ($\epsilon \sim 0.43$ for voids in the \textit{FBI} sample and $\epsilon \sim 0.44$ for voids in the \textit{JP} sample) can be explained by several effects. As previously mentioned in Sect.~\ref{sect:void_size}, the quasi-potential field smooths out small-scale structures and effectively excludes nested sub-voids from the identification process, leaving only the larger embedding underdensities, which have been shown to tend towards sphericity as they expand over time \citep{1984MNRAS.206P...1I}.  Furthermore, our quasi-potential threshold imposed during the watershed procedure implicitly excludes the 'void-in-cloud' population -- voids located within collapsing high-density regions -- which are structures most likely affected by the tidal forces exerted by the surrounding cosmic web and thus could result in less spherical shapes.

The geometric shape of cosmic voids is inherently sensitive to the underlying cosmological model. For instance, \cite{2012MNRAS.426..440B} utilised five sets of cosmological $N$-body simulations to assess how void shapes and sizes vary across different dark energy scenarios relative to $\Lambda$CDM. They found that while the ellipticity distribution of voids shows only marginal variations among the dark energy models, the time evolution of void shapes exhibits distinct signatures when comparing $\Lambda$CDM to quintessence models. Notably, these structural differences are primarily observable in smaller voids ($R_{\rm eff} < 20~h^{-1}\,$Mpc), which we are not assessing in this study.

\subsection{Void recovery}
\label{sect:void_rec}
We identify corresponding voids based on their volume overlap in the two galaxy samples. Using the IoU criterion for the void volumes defined in Sect.~\ref{sec:void_prop_meth}, we recover 425 voids (out of 1074 in the \textit{FBI} sample and 817 in the \textit{JP} sample). The total volume of recovered voids in the \textit{JP} sample occupies $\sim$ 70$\%$ of the potential-thresholded volume, while the total volume of recovered voids in the \textit{FBI} sample occupies $\sim$ 63$\%$. If we consider only the overlapping volume, then the percentage drops to $53\%$.

\begin{figure}[]
    \centering  
    \includegraphics[]{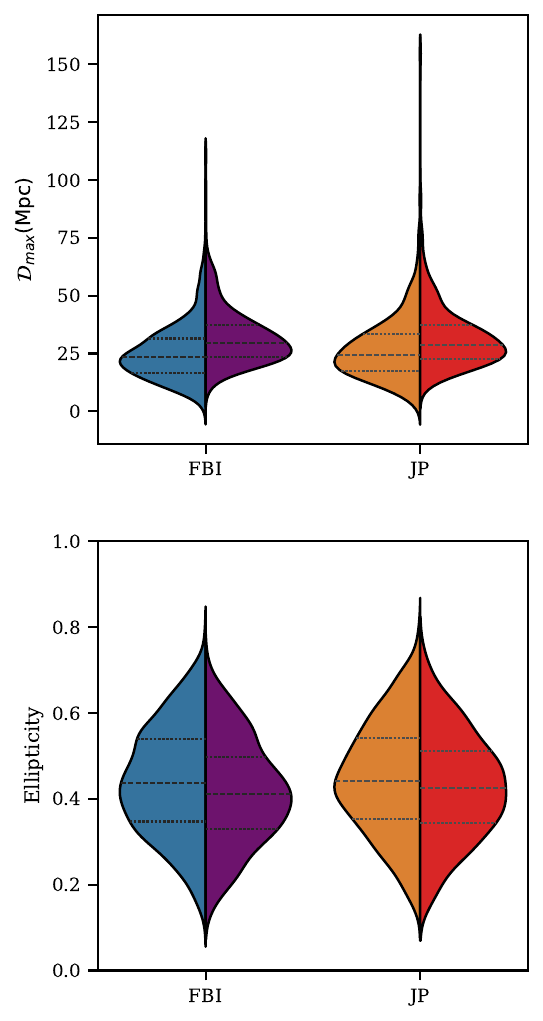}
    \caption{Sizes and shapes of the recovered voids (right halves) compared against the general void population (left halves) in the \textit{FBI} and \textit{JP} galaxy mocks. The top panel shows the maximum boundary distance $D_\mathrm{max}$ while the bottom panel shows the void ellipticity. Internal horizontal lines display the 25th, 50th (median), and 75th percentiles for each distribution.}
    \label{recvoids}
\end{figure}

In Fig.~\ref{recvoids}, we compare the size and shape distributions of recovered voids in the \textit{FBI} and \textit{JP} samples against their respective complete void populations. In the top panel we show that the recovered voids tend to be larger than the general populations; specifically, the median $|D_{\rm max}|$ reaches $29.5$ Mpc in the \textit{FBI} case and $28.5$ Mpc in the \textit{JP} sample. Regarding their shapes, in the bottom panel we show that recovered voids are slightly more spherical, with median ellipticities of $0.41$ and $0.42$.

To comprehensively evaluate the recovery performance of the quasi-potential method, we investigate both the recovered voids sub-population statistics and the object-by-object correspondence. First, we perform separate two-sample KS tests to confirm that the void size and shape distributions are in agreement between the \textit{FBI} and \textit{JP} mocks. For void sizes, using the maximum boundary distance estimator yields a KS statistic of $D_n = 0.0376$ and a $p$-value of $p = 0.9245$. Similarly, a KS test comparing void ellipticities results in $D_n = 0.0706$ and $p = 0.2404$. In both cases, $p > 0.05$, so we fail to reject the null hypothesis. These high $p$-values indicate no statistically significant difference between the \textit{FBI} and \textit{JP} samples regarding either their void size distributions or their ellipticity distributions, supporting the assumption that both populations are drawn from the same underlying distribution.

 Second, to investigate the individual recovered voids' correspondence of sizes and shapes, we calculate the Spearman rank correlation coefficient for both properties. For the maximum boundary distance, we find a remarkably strong correlation of $r_s = 0.94$ ($p < 0.001$). Similarly, for the ellipticity, we find a robust correlation of $r_s = 0.76$ ($p < 0.001$). These highly significant $p$-values confirm a tight physical correspondence between individual recovered voids across both samples.

Furthermore, we assess the relationship between the IoU and maximum boundary distance. In both recovered void samples, we find a positive correlation of $r_s \sim 0.286$ ($p < 0.001$). This suggests a non-random coupling between the two properties, indicative of the fact that larger voids tend to be slightly better recovered, although the modest value of $r_s$ indicates substantial scatter.

\section{Void galaxies in J-PAS mocks}
\label{sec: void galaxies}


A secondary goal of our work is to evaluate the capability of the J-PAS survey to detect expected environmental trends in the properties of massive void galaxies ($M_*$ $\geq$ 10$^{10}$ $M_{\odot}$) such as stellar mass (Fig.~\ref{smf}), star formation activity and colour (Fig.~\ref{gal_prop}) by comparing void core galaxies in the \textit{FBI} and \textit{JP} mocks with galaxies residing in high-density regions (see definitions of the void core and high-density galaxy samples in  Sect.~\ref{sec:voidgals}; see Fig.~\ref{void_core_gal} for an example of void core galaxies within a void). Furthermore, we provide the fraction of void core galaxies that are common to both datasets.

\subsection{Abundance and stellar mass function}

In the \textit{FBI} and \textit{JP} mocks, we identified $37\,224$ and $28\,979$ void core galaxies, respectively, along with $3\,011\,534$ and $2\,012\,043$ galaxies in high-density regions. The lower number of void core galaxies in the \textit{JP} sample is primarily due to the smaller radii of collecting spheres, which result from the artificially increased density in the central regions of voids induced by redshift errors. Conversely, the lower number of galaxies in high-density regions is explained by their scattering into the void interiors.
We matched void core galaxies by their IDs and found 8\,990 common in both galaxy mocks. This represents $24\%$ of the void core galaxies in the \textit{FBI} and 31$\%$ in the \textit{JP} samples, respectively. Our subsequent analysis of environmental trends focuses on the full individual samples rather than the overlapping subset to maximise statistical significance.

\begin{figure} [t]
    \centering
    \includegraphics[]{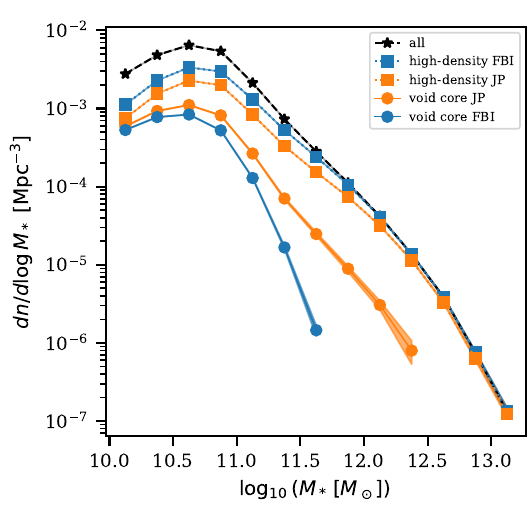}
    \caption{Stellar mass function for void core galaxies (solid lines, circles), galaxies in high-density regions (dotted lines, squares)  and the overall galaxy population (dashed line, stars). The thin, shaded regions represent the 25th and 75th percentiles of the median values obtained from 1,000 bootstrap resamples.}
    \label{smf}
\end{figure}

\begin{figure*}[t]
    \centering  
    \begin{minipage}[t]{\textwidth}
    \centering  
    \includegraphics[]{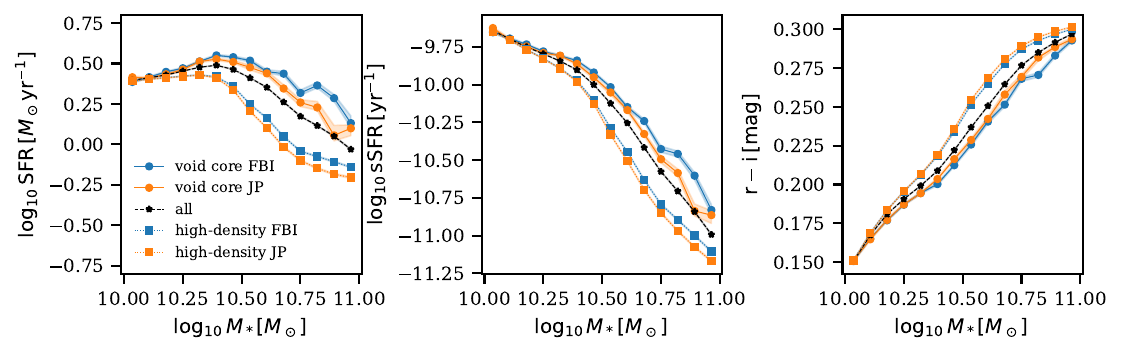}
    \caption{Star formation rate (left), specific star formation rate (centre) and colour (right) of void core (solid lines, circles), high-density galaxies (dotted lines, squares) in the \textit{FBI} and \textit{JP} samples and the population of all galaxies (dashed line, stars), for reference. Intervals correspond to the 25th and 75th percentiles of 1000 bootstrapped median distribution}
    \label{gal_prop}
    \end{minipage}
    \hfill
\end{figure*}

 As voids are underdense regions, they are expected to host a population of lower-mass galaxies compared to denser regions. Fig.~\ref{smf} compares the SMFs for galaxies residing in void cores, high-density regions, and the overall galaxy population. Filled intervals represent the 25th and 75th percentiles of 1\,000 bootstrap resamples. Due to the lower number density, the SMF of void core galaxies exhibits lower values in both the \textit{FBI} and \textit{JP} samples. While the shapes of the distributions are consistent at the low-mass end ($M_* < 10^{10.7}\,M_{\odot}$), a significant divergence occurs at $M_* \geq 10^{10.7}\,M_{\odot}$. This suggests a lack of massive galaxies in void interiors, with the \textit{FBI} sample's SMF dropping sharply at $M_* \sim 10^{11.5}\,M_{\odot}$. Although the \textit{JP} sample's SMF shows a shift toward higher stellar masses - likely due to contamination of massive galaxies from overdense regions - it remains distinct from the SMF of the corresponding galaxies in high-density regions. This implies that the low-mass signature of the void environment is preserved despite the effect of photometric redshift errors.


\subsection{Colour and star formation activity}

To isolate the environmental influence of voids from the dependence of galaxy properties on stellar mass, we compared our galaxy samples by binning in stellar mass. Furthermore, we restrict our analysis of galaxy properties to the stellar mass range $10^{10} \leq M_\star \leq 10^{11}\,M_\odot$. This is because for values higher than $10^{11} M_*$, an upturn in the SFR can be observed. The upturn is caused by AGN feedback, which is insufficient to keep most massive galaxies quenched \citep{2023MNRAS.526.4978S, 2025MNRAS.543.2204L}. An additional reason for limiting the mass range is given by the low number of void core galaxies with masses above  $10^{11} M_*$.

In Fig.~\ref{gal_prop}, we show how the galaxy properties of void core, high-density regions and the overall galaxy populations vary with their stellar mass. Intervals denote the interquartile range of 1\,000 bootstrap realisations of the median.
Regardless of the galaxy sample, we notice a decrease in star formation activity with stellar mass. This can be attributed to the increase of passive fraction of galaxies due to the efficiency of AGN feedback \citep{2023MNRAS.526.4978S, 2025MNRAS.543.2204L}. Void core galaxies in both the \textit{FBI} and \textit{JP} samples exhibit consistent properties, characterised by slightly elevated SFR, sSFR, and bluer colours compared to both the high-density counterparts and the general galaxy population. This suggests that, despite the photometric redshift errors which contaminate void interiors with massive galaxies, we can expect to capture the general trend previously observed in void galaxies.


The effect of the environment on galaxy properties has been extensively studied in observations. Several studies showed how void galaxies tend to have lower stellar masses \citep{Kreckel_2012, 2021ApJ...906...97F, 2022A&A...668A..69E}, bluer colours \citep{Grogin_1999, Kreckel_2012, 2012MNRAS.426.3041H, 2021ApJ...906...97F, Rodr_guez_Medrano_2023}, higher SFR \citep{Kreckel_2012, 2017MNRAS.464..666B, 2021ApJ...906...97F, Rodr_guez_Medrano_2023, 2025A&A...698A.196Z} with respect to their counterparts in denser regions. In particular, our predictions are consistent with those of \citet{Rodr_guez_Medrano_2023}, who found that void galaxies in the SDSS-DR16 are slightly bluer ($g-r$) than the general population within the stellar mass range $10^{10}\text{--}10^{11}\,M_{\odot}$. For these same stellar masses, they observed that while both populations follow similar declining SFR trends, void galaxies consistently exhibit marginally higher SFRs at fixed stellar mass.

The physical properties of void galaxies have also been characterised using 
various hydrodynamical $\Lambda$CDM simulations. In particular, \citet{2022MNRAS.517..712R} explored the properties of central galaxies with stellar mass ranges between $10^{9}$ and $10^{11}\,M_{\odot}$ within the EAGLE hydrodynamic simulation. By partitioning their sample based on the distance from the void centre -- into inner void, outer void, wall, and skeleton regions -- they found that galaxies in the outer void regions possess the highest sSFR at a fixed stellar mass. This seems to contradict our results, although the authors noted that these results remain subject to significant statistical uncertainties.

An earlier study using the Horizon-AGN simulation by \citet{2020MNRAS.493..899H} revealed that, as a function of distance from the void centre, galaxies exhibit increasing sSFR and an overall decrease of SFR. Furthermore, the authors report that low-mass galaxies are more abundant than their massive counterparts close to the void centre.

\citet{2024ApJ...962...58C} utilised the IllustrisTNG300 suite 
to demonstrate that void galaxies are, on average, younger, bluer, and less 
massive than non-void populations. Their findings quantify this enhancement 
in star formation, noting a sSFR approximately 20\% higher than that observed 
in the general field population. Another recent study, which used IllustrisTNG-300 \citep{2026arXiv260418209H} investigated the environmental effect of different components of the cosmic web, taking into account its multiscale hierarchical nature. The authors found that void galaxies are bluest, presenting the lowest metallicities and retaining the highest fraction of baryons while actively star-forming.

\section{Conclusions}
\label{sec:conc}

Photometric surveys can excel in mapping the large-scale structure of the Universe. Their ability to perform rapid, multi-band imaging allows for the acquisition of large datasets spanning thousands of square degrees, far more quickly than spectroscopic campaigns and with fewer selection effects.



Cosmic voids are sensitive probes of the large-scale structure and offer pristine environments for the study of galaxy properties, but their identification in photometric surveys is complicated by the large photometric redshift errors. Using FLAMINGO (L1$\_$m8) hydrodynamic simulation at redshift snapshot of $z=0.3$, we constructed two J-PAS galaxy mocks with stellar masses in the range $10^{10} \leq M_\star \leq 10^{13.2}\,M_\odot$: a FLAMINGO-based ideal (\textit{FBI}), with an apparent magnitude limit $m_{i} < 20$ and a FLAMINGO-based \textit{JP}, where we additionally modelled realistic photometric redshift errors. We demonstrated that an approach based on a quasi-gravitational potential field (henceforth quasi-potential), computed using the logarithm of the galaxy number density, provides a robust alternative to the density field for identifying and characterising dynamically relevant voids. The inherent smoothing capability of the quasi-potential effectively mitigates photometric redshift uncertainties, ensuring good agreement between the identified void properties. Furthermore, to validate the environmental impact of our voids, we demonstrate that our framework successfully reproduces the expected physical trends in the properties of the massive void-galaxy population within the J-PAS mocks, such as lower stellar masses and enhanced star formation activity.

The following results highlight the performance of our quasi-potential method of void and void galaxies identification when subjected to realistic J-PAS photometric redshift uncertainties:

(i) We found comparable void abundances at z = 0.3 in the quasi-potential thresholded simulation volume of the \textit{FBI} and \textit{JP} galaxy mocks. Additionally, we recover $\sim 40\%$ of individual voids in the \textit{FBI} sample from the \textit{JP} sample, which account for $63\%$ of the thresholded simulation volume in the \textit{FBI} sample.

(ii) We find that the size and shape distributions of the full void populations in the \textit{JP} and \textit{FBI} mocks are in good agreement despite the impact of redshift uncertainties (Figs.~\ref{sizes}, \ref{shapes}). For the subset of individual recovered voids, we find a high degree of object-by-object consistency for the maximum boundary distance size estimator and for the ellipticity. Furthermore, regardless of the galaxy mock used, these recovered voids tend to be systematically larger and more spherical in comparison with the general void population (Fig.~\ref{recvoids}).

(iii) Analysis of the spherical and boundary density profiles (Fig.~\ref{sphprof}, \ref{bdprof}) reveals that the voids in the \textit{JP} sample present slightly lower density watershed boundaries and denser central regions. This trend comes as a consequence of the photometric redshift errors, which cause the displacement of galaxies from high-density regions to scatter within the void interiors and to produce an artificial increase in density profiles.


(iv) The SMF of void core galaxies in the \textit{FBI} sample reveals an overall lower abundance of galaxies and a lack of more massive galaxies ($M_* > 10^{11.5} M_{\odot}$), when compared to galaxies located in high-density regions (Fig.~\ref{smf}). The SMF for void core galaxies in the \textit{JP} sample is in good agreement with the one in the \textit{FBI} sample in the stellar mass range of $10^{10}-10^{11}$ $\,M_{\odot}$. However, a contamination of \textit{JP} sample voids with more massive galaxies is clearly present past $\simeq10^{11}$$\,M_{\odot}$ when we can observe a divergence of the two SMFs.

(v) We found slightly higher values in the SFR and sSFR, alongside bluer colours at fixed stellar mass, for void core galaxies within both \textit{FBI} and \textit{JP} samples when compared to galaxies located in high-density regions (Fig.~\ref{gal_prop}).

An important limitation that needs to be restated is the high stellar mass ($\geq$ 10$^{10}$ $M_\odot$) of the galaxies in our sample. Most void galaxy studies focus on low mass galaxies, typically $10^{8} - 10^{9} M_{\odot}$, which are more abundant in voids. Since FLAMINGO simulation does not provide galaxy properties such as luminosities and star formation rates for those masses, our void galaxies are on the heavier side and in lower number. This might reduce the real difference in star formation activity and stellar mass between void galaxies and galaxies in high-density regions.  We also would like to remind that the current study represents an application of the quasi-potential method to J-PAS galaxy mocks, rather than a methodological validation paper. A quantitative calibration of the optimal parameters (including grid resolution, log-transforms, and quasi-potential thresholds) is the core subject of a future paper (McCarthy et al., in prep.).

In FLAMINGO-based J-PAS galaxy mocks with $m_i < 20$ at $z=0.3$, a quasi-potential has proven to effectively mitigate photometric redshift uncertainties, providing a robust framework for identifying dynamically relevant voids and characterising their internal massive galaxy populations. Our analysis suggests that this framework might enable the robust detection of voids and the preservation of massive void galaxy trends in forthcoming J-PAS data. Further validation is still needed using realistic J-PAS features such as selection functions, full photo-z posteriors, survey masks and redshift-space distortions. Furthermore, extending this method to larger-volume surveys, such as Euclid, will provide the statistical power necessary to further characterise the influence of the void environment on galaxy formation and evolution. Finally, this quasi-potential based method of void identification offers an attractive pathway toward tighter cosmological constraints in analyses that rely on stacking, such as the Alcock–Paczynski (AP) test, the Integrated Sachs-Wolfe (ISW) effect, and weak lensing.

\begin{acknowledgements}
We acknowledge the support by the Estonian Ministry of Education and Research (grant TK202), Estonian Research Council grant (PRG2172, PRG1006, PSG700, PRG2159, PRG3034) and the European Union's Horizon Europe research and innovation programme (EXCOSM, grant No. 101159513). This work is based on observations made with the JST/T250 telescope and JPCam at the Observatorio Astrofísico de Javalambre (OAJ), in Teruel, owned, managed, and operated by the Centro de Estudios de Física del Cosmos de Aragón (CEFCA). We acknowledge the OAJ Data Processing and Archiving Department (DPAD) for reducing and calibrating the OAJ data used in this work. Funding for the J-PAS Project has been provided by the Governments of Spain and Aragón through the Fondo de Inversión de Teruel, European FEDER funding and the Spanish Ministry of Science, Innovation and Universities, and by the Brazilian agencies FINEP, FAPESP, FAPERJ and by the National Observatory of Brazil with additional funding also provided by the University of Tartu and by the J-PAS Chinese Astronomical Consortium. VM thanks CNPq (Brazil), CAPES (Brazil) and FAPES (Brazil) for partial financial support. 
\end{acknowledgements}

%


\bibliographystyle{aa} 
\bibliography{ref} 

\end{document}